\documentclass[a4paper,11pt]{article}
\pdfoutput=1
\usepackage{jcappub}
\usepackage[T1]{fontenc}
\usepackage{subfig}
\usepackage{caption}
\usepackage{hyperref}
\usepackage{amsmath}
\usepackage{mathtools}
\usepackage{tablefootnote}
\usepackage{soul}
\usepackage{ulem}
\usepackage{color}
\usepackage{float}
\usepackage{makecell}
\usepackage[usenames,dvipsnames,svgnames,table]{xcolor}

\definecolor{darkred}{RGB}{175,0,0}

\title{GW$\times$LSS: Chasing the Progenitors of Merging Binary Black Holes} 

\author[a,b]{Giulio Scelfo,}
\emailAdd{giulio.scelfo@studenti.unipd.it}

\author[b,c]{Nicola Bellomo,}
\emailAdd{nicola.bellomo@icc.ub.edu}

\author[b]{Alvise Raccanelli,}
\emailAdd{alvise@icc.ub.edu}

\author[a,d,e,f]{Sabino Matarrese,}
\emailAdd{sabino.matarrese@pd.infn.it}

\author[b,g]{Licia Verde}
\emailAdd{liciaverde@icc.ub.edu}

\affiliation[a]{Dipartimento di Fisica e Astronomia G. Galilei, Universit\`a degli Studi di Padova, via Marzolo 8, I-35131 Padova, Italy.}
\affiliation[b]{ICC, University of Barcelona, IEEC-UB, Mart\'i  i Franqu\`es, 1, E-08028 Barcelona, Spain}
\affiliation[c]{Dept. de  F\'isica Qu\`antica i Astrof\'isica, Universitat de Barcelona, Mart\'i  i Franqu\`es 1, E-08028 Barcelona, Spain}
\affiliation[d]{INFN - Sezione di Padova, via F. Marzolo 8, I-35131 Padova, Italy.}  
\affiliation[e]{INAF - Osservatorio Astronomico di Padova, vicolo dell'Osservatorio 5, I-35122 Padova, Italy.}  
\affiliation[f]{Gran Sasso Science Institute, viale F. Crispi 7, I-67100 L'Aquila, Italy.}
\affiliation[g]{ICREA, Pg. Lluis Companys 23, Barcelona, E-08010, Spain.}

\abstract{
Are the stellar-mass merging binary black holes, recently detected by their gravitational wave signal, of stellar or primordial origin? Answering this question will have profound implications for our understanding of the Universe, including the nature of dark matter, the early Universe and stellar evolution.
We build on the idea that the clustering properties of merging binary black holes can provide information about binary formation mechanisms and origin.
The cross-correlation of galaxy with gravitational wave catalogues carries information about whether black hole mergers trace more closely the distribution of dark matter -- indicative of primordial origin -- or that of stars harboured in luminous and massive galaxies -- indicative of a stellar origin. 
We forecast the detectability of such signal for several forthcoming and future gravitational wave interferometers and galaxy surveys, including, for the first time in such analyses, an accurate modelling for the different merger rates, lensing magnification and other general relativistic effects.
Our results show that forthcoming experiments could allow us to test most of the parameter space of the still viable models investigated, and shed more light on the issue of binary black hole origin and evolution.
}
 
\begin{document}
\maketitle

\section{Introduction}
\label{sec:intro}
The first detection of gravitational waves (GWs) emitted by the coalescence of two black holes (BHs) of approximately $30\ M_{\odot}$ \cite{abbott:firstligodetection, abbott:firstligodetectionproperties} opened the era of gravitational waves astronomy, not only by confirming General Relativity predictions, but also establishing a new way to observe and analyse the cosmos. Even if some authors expected such massive progenitors to be the first sources to be detected, see e.g., Refs. \cite{belczynski:massivebhsmergers, dominik:massivebhsmergersI, dominik:massivebhsmergersII, dominik:massivebhsmergersIII}, this fact was hailed by part of the community as unexpected and led some researchers to suggest that such events may not be uncommon. Indeed other GWs events followed \cite{abbott:secondligodetection, abbott:thirdligodetection, abbott:fourthligodetection, abbott:fifthligodetection} and confirmed that apparently a significant fraction of the detected progenitors has masses between $20$ and $40\ M_{\odot}$. Such large masses of the progenitors are not incompatible with classical stellar/binary evolution~\cite{belczynski:highmassbhI, belczynski:highmassbhII, belczynski:highmassbhIII}. Nevertheless the possibility that BHs with an origin different from the standard end-point of stellar evolution and constituting a significant fraction of the dark matter regained interest~\cite{bird:pbhasdarkmatter, clesse:pbhmerging}.

The authors of Refs. \cite{Zeldovich:pbh, Hawking:pbh1971, CarrHawking:pbh} were the first to show analytically that, because of large density fluctuations in the primordial cosmic fluid, some extremely overdense regions in the primordial Universe may have overcome pressure forces and have collapsed to give birth to so-called Primordial Black Holes (PBHs). These results were later confirmed by the authors of Ref. \cite{musco:pbh}, who were the first to provide  general relativistic numerical computations of PBHs formation during the radiation-dominated era. Even if the PBHs formation mechanism is unknown, many proposals have been made, including collapse of cosmic string loops \cite{Polnarev88, HAWKING1989237, WICHOSKI1998191} and domain walls \cite{BEREZIN198391, Ipser84}, bubble collisions \cite{Crawford82, LA1989375}, through the collapse of large fluctuations produced during inflation as pioneered in Refs. \cite{ivanov:pbhfrominflationI, ivanov:pbhfrominflationII, Bellido:pbh} or even through the collapse of (interacting) dark matter clumps \cite{shandera:pbhfromdarkmatter}. 

Given the high interest in PBHs as dark matter candidates, a remarkable amount of different observational constraints have been obtained, including constraints coming from gravitational lensing effects \cite{barnacka:femtolensingconstraint, katz:femtolensingconstraint, griest:keplerconstraint, niikura:microlensingconstraint, tisserand:microlensingconstraint, calchinovati:microlensingconstraint, alcock:microlensingconstraint, mediavilla:microlensingconstraint, wilkinson:millilensingconstraint, zumalacarregui:supernovaconstraint}, dynamical effects \cite{graham:whitedwarfconstraint, capela:neutronstarconstaint, quinn:widebinaryconstraint, brandt:ufdgconstraint, alihaimoud:pbhmergerrate, magee:mergerrate} and accretion effects \cite{gaggero:accretionconstraints, ricotti:cmbconstraint, alihamoud:pbhaccretion, poulin:cmbconstraint, bernal:cmbconstraint}. Even if these constraints cover the whole mass range and seem to disfavour PBHs as a significant fraction of the dark matter, these results are far from being conclusive due to the variety of assumptions involved, see e.g., \cite{KfirBlum, bellomo:emdconstraints}. Some mass ``windows'' still exist, for instance one around $10^{-10}\ M_\odot$ and another $10\ M_\odot$, where the latter one can be probed by future GWs observatories as Advanced LIGO (aLIGO) \cite{ALIGO:aligo} or Einstein Telescope (ET) \cite{Sathyaprakash:ET}.

Despite the fact PBHs may not constitute the totality of the dark matter, it is valuable to explore different ways to determine if mergers progenitors' origin is stellar or primordial. Several proposals have been made, including testing the cross-correlation between galaxy and GW maps \cite{raccanelli:pbhprogenitors, raccanelli:gwastronomy, nishikawa:pbhmergers}, BHs binaries orbital eccentricity \cite{cholis:orbitaleccentricities}, fast radio bursts \cite{munoz:fastradioburst}, the BHs mass function \cite{kovetz:pbhmassfunction, kovetz:pbhandgw}.

In this work we focus on developing further the cross-correlation approach suggested by the authors of Ref. \cite{raccanelli:pbhprogenitors}, who show that the  statistical properties of the type of galaxy (or halo) hosting a GWs event can provide information about the system origin (stellar or primordial). In fact, in more massive halos the typical velocities are higher than those in the less massive ones (the reader can think of the virial theorem or check numerical simulations \cite{Munari:velocity_mass_halos}). As a consequence, it is much more probable that two PBHs form a gravitationally bound binary through GWs emission in low-mass halos, since the cross section of such process is inversely proportional to some power of the relative velocity of the progenitors. The higher velocity dispersion of high-mass halos make this process for PBHs less likely to happen. In addition, low-mass halos tend to be less luminous \cite{Vale:luminosity_mass_halos} than high-mass ones and trace more closely the dark matter distribution than high-mass halos. On the other hand  the merger probability for stellar black holes is more likely to correlate with galaxies' (or halos') stellar mass, hence stellar black holes mergers tend to happen in more luminous and massive halos. Recall that star formation efficiency increases with halo mass for halo of masses below $10^{12} M_{\odot}$. It decrease for higher mass-halos but these are very rare and more closely associated to galaxy clusters rather than galaxies. We refer the interested reader to Ref. \cite{erb:starformationrate} and references therein.

Therefore, once a significant number of GWs coming from BHs mergers will be detected, it will be useful to correlate the corresponding events map with a map of galaxies. If the BHs progenitors were mostly of stellar origin, GWs events would be associated with massive halos, and thus would be highly correlated with luminous galaxies. On the other hand, if these progenitors were more likely to have primordial origin, GWs would come mostly from low mass halos (i.e., they would be poorly correlated with luminous galaxies). While mergers of BHs of primordial origin tend to trace the filaments (dark-matter/low mass-halos distribution\footnote{Since we are performing a statistical analysis, the presence of subhalos inside high mass halos does not affect significantly the results because of their relative abundance compared to ``field'' halos (located in lower density regions such as filaments) of the same mass.}) of the large-scale structure, stellar-BHs mergers tend to trace the distribution of galaxies of high stellar mass. The clustering properties of these two populations and the statistical properties the two maps are different. Low-mass halos tracing filaments are less strongly clustered than high (stellar) mass galaxies: in particular they have different \textit{bias} parameters. The bias parameter governs the ratio of clustering amplitude of the selected tracer to that of the dark matter. 

At the moment too few GWs events have been detected to measure the auto and cross correlation of maps of GWs events and galaxies, but during next LIGO's runs, thousands of events are likely to be detected due to the improved sensitivity. On the other hand, during the next decade a large volume of the Universe at high redshift will be surveyed thanks to several surveys, as EMU \cite{Norris:EMU}, DESI \cite{Aghamousa:desi} or SKA \cite{maartens:SKA}, which we consider in the rest of the paper. Here we develop a Fisher and a $\Delta\chi^2$ analysis to forecast the ability of future surveys to accomplish this goal. We improve the treatment of Ref. \cite{raccanelli:pbhprogenitors} in different ways. Firstly, we consider both cross and auto correlation terms between our tracers (more details in section~\ref{sec:ggwcorrelation}), whereas the latter ones were previously neglected. Secondly, our computations include all possible general relativistic effects which, as we will show, can influence the results. Thirdly, we use a theoretically-motivated PBHs merger rate without neglecting its redshift dependence. Finally, when modelling the GWs events distribution, we provide for the first time an analytic expression for the magnification bias of gravitational waves. This is a step forward in the study of the lensing of gravitational waves coming from black holes or neutron stars mergers and can provide significant insight on BHs binary formation and evolution~\cite{dai:gravitationallensing, oguri:gravitationallensing, wang:lensing}, on estimates of the luminosity distance \cite{bertacca:luminositydistance} or even on alternatives to General Relativity~\cite{camera:gwlensing}.

The paper is structured as follows: in section \ref{sec:ggwcorrelation} we explain the methodology used in this work and introduce the multi-tracer cross-correlation formalism, in section \ref{sec:tracers} we characterize the galaxies (\ref{subsec:galaxies}) and GWs (\ref{subsec:gw}) tracers considered while in section \ref{sec:results} we present the results of the forecast. Finally we conclude in section \ref{sec:conclusions}.


\section{Methodology and Galaxy--GW Correlation}
\label{sec:ggwcorrelation}
Since BH-BH mergers do not have an electromagnetic counterpart, the identification of their host object is impossible even if the event is measured by more than three detectors. Because of the poor localisation in the sky of the GWs events, the GWs maps are typically very ``low resolution''. For this reason we approach the problem in a statistical way, by using measurements and statistical properties of their number counts. In particular, we work in harmonic space and we consider the number counts angular power spectrum, $C_\ell$, where only low multipoles $\ell$ are considered because of the maps' low angular resolution. The maximum multipole $\ell_\mathrm{max}$ is determined by the  angular resolution $\theta$ that can be achieved: $\ell_\mathrm{max}\sim 180^o/\theta$. For the aLIGO+Virgo network $\ell_\mathrm{max}=20$, once also LIGO-India and KAGRA are included, we improve the spatial resolution up to $\ell_\mathrm{max}=50$ and finally with the futuristic Einstein Telescope,  $\ell_\mathrm{max}=100$ will be reached. The interested reader can check Refs. \cite{schutz:lmax, klimenko:lmax, sidery:lmax, namikawa:lmax}. We discuss the benefits of having higher resolution in GWs maps in section \ref{subsec:forecasts}.
 
In the following we assume to have (tomographic) maps of GWs events and of galaxies (i.e., the {\it tracers}). The observed harmonic coefficients  used to compute the angular power spectra are given by 
\begin{equation}
a_{\ell m}^X(z_i) = s_{\ell m}^X(z_i) + n^X_{\ell m}(z_i),
\end{equation}
where $s^X_{\ell m}$ and $n^X_{\ell m}$ are the partial wave coefficients of the signal and of the noise for tracer $X$. We consider the noise angular power spectrum to be given only by a shot noise term $\mathcal{N}^X_\ell(z_i)$ and we assume that the noise terms from different experiments and different redshift bins are uncorrelated, which means that
\begin{equation}
\langle n^X_{\ell m}(z_i) n^{Y^*}_{\ell' m'}(z_j) \rangle = \delta_{\ell\ell'} \delta_{mm'} \delta_{XY} \delta_{ij} \mathcal{N}^X_\ell(z_i),
\label{noise}
\end{equation}
where $\delta$ denotes the Kronecker delta. The expectation value of the signal gives the $C_\ell$s \cite{raccanelli:crosscorrelation, pullen:crosscorrelation},
\begin{equation}
\langle s^X_{\ell m}(z_i) s^{Y^*}_{\ell' m'}(z_j) \rangle = \delta_{\ell\ell'} \delta_{mm'} C^{XY}_\ell(z_i,z_j),
\end{equation} 
while the signal-cross-noise expectation value is given by 
\begin{equation}
\langle s^X_{\ell m}(z_i) n^{Y^*}_{\ell' m'}(z_j) \rangle = 0,
\end{equation}
since we assume signal and noise  to be statistically independent. Finally, the observed angular power spectrum $\tilde{C}^{XY}_\ell(z_i,z_j)$ reads as
\begin{equation}
\langle a^X_{\ell m}(z_i) a^{Y^*}_{\ell'm'}(z_i)\rangle = \delta_{\ell \ell'} \delta_{mm'}\tilde{C}^{XY}_\ell(z_i,z_j) = \delta_{\ell \ell'} \delta_{mm'} \left[C^{XY}_\ell(z_i,z_j) + \delta_{XY} \delta_{ij} \mathcal{N}^X_\ell(z_i) \right].
\label{eq:tilde_Cls}
\end{equation}

Following the notation of Ref. \cite{bonvin:cl} (which is re-arranged differently than the standard way, but reflects how the public code \texttt{CLASS} \cite{blas:class, didio:classgal} is structured) and generalizing their formalism to the case of multiple tracers we can write the angular power spectrum as
\begin{equation}
C_{\ell}^{XY}(z_i,z_j) = \frac{2}{\pi} \int\frac{dk}{k} \mathcal{P}(k) \Delta^{X,z_i}_{\ell}(k) \Delta^{Y,z_j}_{\ell}(k),
\label{eq:Cls}
\end{equation}
where $\{X,Y\}$ stands for the different tracers (galaxies and GWs in our case), $\mathcal{P}(k)= k^3P(k)$ is the primordial  power spectrum and
\begin{equation}
\Delta^{X,z_i}_{\ell}(k) = \int_{z_i-\Delta z}^{z_i+\Delta z} dz \frac{dN_X}{dz}W(z,z_i)\Delta^X_\ell(k,z),
\label{eq:Delta_l}
\end{equation}
where we have introduced a window function\footnote{In this work we use a Top-Hat window function.} $W(z,z_i)$, centered at redshift $z_i$ with bin half-width $\Delta z$, the source number density per redshift interval $\frac{dN_X}{dz}$, and the tracer $X$ angular number count fluctuation $\Delta^X_\ell(k,z)$. The integral of $\displaystyle W(z,z_i)\frac{dN_X}{dz}$ is normalized to unity. In general the observed number count fluctuation receives contributions from density ($\mathrm{den}$), velocity ($\mathrm{vel}$), lensing ($\mathrm{len}$) and gravity ($\mathrm{gr}$) effects \cite{bonvin:cl, challinor:deltag}:
\begin{equation}
\Delta_\ell(k,z) = \Delta^\mathrm{den}_{\ell}(k,z) + \Delta^\mathrm{vel}_\ell(k,z) + \Delta^\mathrm{len}_\ell(k,z) + \Delta^\mathrm{gr}_\ell(k,z).
\label{eq:numbercount_fluctuation}
\end{equation}
We report in appendix \ref{app:relativistic_number_counts} the complete form of the various terms in equation \eqref{eq:numbercount_fluctuation}. Even if the bias parameter $b_X$ of the tracer $X$ enters only in the density contribution $\Delta^\mathrm{den}_{\ell}(k,z)$, we cannot overlook the effect of the other terms on the signal-to-noise, as sometimes done in the literature. Since the public code \texttt{CLASS} allows us to choose whether include or not the velocity, lensing and gravity effects in the computation of the $C_\ell$, we estimate the error one would introduce by neglecting these contributions in section \ref{subsec:greffects}. The reader interested in a more in general discussion on the importance of a correct modelling of an observable can check Refs.~\cite{bellomo:multiclassI, bernal:multiclassII}. To compute the angular power spectra we extend the public code \texttt{CLASS} to include the possibility to have different tracers ($X\neq Y$). We present this new version of \texttt{CLASS}, called \texttt{Multi\_CLASS},\footnote{Users can find and download the code on the GitHub page \url{https://github.com/nbellomo/Multi_CLASS}.} in Refs.~\cite{bellomo:multiclassI, bernal:multiclassII}. It should be noticed that, in the case of different tracers, the angular projections are not symmetrical under the exchange of redshift $z_i\longleftrightarrow z_j$, therefore, if we have $n$ redshift bins,  we have to compute $n^2$ different $C_\ell$, while for identical tracers we have to compute only $n(n+1)/2$ angular power spectra.

Hereafter we consider two tracers, galaxies and gravitational waves, labelled by $\mathrm{\{g,GW\}}$. For illustrative purposes, the tracers have been divided in three redshift bins with central  redshift values $\{z_1,\ z_2,\ z_3\}=\{1.5,\ 2.5,\ 3.5\}$ and bin half-width $\Delta z = 0.5$. This choice refers to the main generic case discussed in section \ref{subsec:greffects}; in section \ref{subsec:forecasts}, we also provide results for specific surveys:  EMU~\cite{Norris:EMU} (a wide-field radio continuum survey planned for the new Australian Square Kilometre Array Pathfinder telescope), DESI \cite{Aghamousa:desi} (a Stage IV ground-based dark energy experiment planned to study baryon acoustic oscillations and the growth of structure through redshift-space distortions with a wide-area galaxy and quasar redshift survey), and SKA~\cite{maartens:SKA} (as a radio continuum survey with $5\ \mu Jy$ flux limit at redshift $z<5$.).
For gravitational waves experiments we consider aLIGO \cite{ALIGO:aligo} (a GWs experiment which currently being developed with enough sensitivity to detect $30\ M_{\odot}$ binary black holes mergers up a redhsift $z_{\mathrm{max}}=0.4$), LIGO-India~\cite{unnikhrishnan:indigo} and KAGRA \cite{somiya:kagra} (we also include these last two detectors as to improve event localization in the sky and thus increase the resolution of the resulting GW map) and Einstein Telescope (ET) \cite{Sathyaprakash:ET} (a planned GWs detector with higher sensitivity and resolution than aLIGO).

We estimate the capability of future surveys to determine BHs mergers progenitors' origin in two different ways, one more conservative that follows the approach suggested in Ref.~\cite{raccanelli:pbhprogenitors}, the other more optimal and closer to an actual data analysis but that relies on modelling well some properties of the tracers that are currently still uncertain. We perform what can be seen as a null hypothesis testing, comparing two models, one in which progenitors origin is stellar, the other in which is primordial. We assume one model as fiducial and we check if the alternative model can be differentiated from the fiducial one by computing a Signal-to-Noise ratio $S/N$. The null hypothesis is that the model is indistinguishable from the fiducial, which happens for low values of the Signal-to-Noise ratio ($S/N\lesssim 1$).

The first procedure relies on a standard Fisher analysis, where we consider a parameter set $\{\theta_\alpha\}$, given by the cold dark matter physical density $\omega_\mathrm{cdm}$, the baryon physical density $\omega_{\mathrm{b}}$, the angular scale of the sound horizon at decoupling $100\theta_s$, the amplitude of scalar perturbations $\log 10^{10}A_s$, the spectral index $n_s$ and an effective bias $B_\mathrm{g}$ and $B_\mathrm{GW}$ of galaxies and GWs, respectively\footnote{The fiducial values of the five standard cosmological parameters reads as
\begin{equation}
\{\omega_\mathrm{cdm}, \omega_\mathrm{b}, 100\theta_s, \log 10^{10}A_s, n_s\} = \{0.12038, 0.022032, 1.042143, 3.0980, 0.9619 \}.
\end{equation}
The fiducial values of the effective galaxy and GWs bias, $B_{\mathrm{g}}$ and $B_{\mathrm{GW}}$, are discussed in detail in section \ref{sec:tracers}, but see also table \ref{tab:effective_bias}, where results discussed in section \ref{sec:tracers} are summarized.}. More details on how we calculate the effective bias of the tracers are given in section \ref{sec:tracers}. Following the authors of Refs. \cite{Bunn:Fisher, Tegmark:Fisher, Vogeley:FIsher}, we write the upper triangular part of the covariance matrix $\mathcal{C}_\ell$ as
\begin{equation}
\mathcal{C}_\ell=
\begin{bmatrix}
\tilde{C_\ell}^{\mathrm{gg}}(z_1,z_1) & \tilde{C_\ell}^{\mathrm{gg}}(z_1,z_2) & \tilde{C_\ell}^{\mathrm{gg}}(z_1,z_3) & \tilde{C_\ell}^{\mathrm{gGW}}(z_1,z_1) & \tilde{C_\ell}^{\mathrm{gGW}}(z_1,z_2) & \tilde{C_\ell}^{\mathrm{gGW}}(z_1,z_3)\\
 & \tilde{C_\ell}^{\mathrm{gg}}(z_2,z_2) & \tilde{C_\ell}^{\mathrm{gg}}(z_2,z_3) &\tilde{C_\ell}^{\mathrm{gGW}}(z_2,z_1) & \tilde{C_\ell}^{\mathrm{gGW}}(z_2,z_2) & \tilde{C_\ell}^{\mathrm{gGW}}(z_2,z_3) \\
 &  & \tilde{C_\ell}^{\mathrm{gg}}(z_3,z_3) & \tilde{C_\ell}^{\mathrm{gGW}}(z_3,z_1) & \tilde{C_\ell}^{\mathrm{gGW}}(z_3,z_2) & \tilde{C_\ell}^{\mathrm{gGW}}(z_3,z_3)\\
 &  &  & \tilde{C_\ell}^{\mathrm{GWGW}}(z_1,z_1)& \tilde{C_\ell}^{\mathrm{GWGW}}(z_1,z_2) & \tilde{C_\ell}^{\mathrm{GWGW}}(z_1,z_3)\\
 &  &  &  & \tilde{C_\ell}^{\mathrm{GWGW}}(z_2,z_2) & \tilde{C_\ell}^{\mathrm{GWGW}}(z_2,z_3)\\
 &  &  &  &  & \tilde{C_\ell}^{\mathrm{GWGW}}(z_3,z_3)\\
\end{bmatrix},
\label{eq:covariance_matrix_first_method}
\end{equation}
where the lower triangular part can be easily obtained noticing that the covariance matrix is symmetric. The covariance matrix is then used to compute the Fisher matrix elements as
\begin{equation}
F_{\alpha \beta}=f_\mathrm{sky}\sum_\ell \frac{2\ell+1}{2} \mathrm{Tr}\left[\mathcal{C_\ell}^{-1} (\partial_\alpha \mathcal{C_\ell})\mathcal{C_\ell}^{-1}(\partial_\beta \mathcal{C_\ell})\right],
\label{eq:Fisher}
\end{equation}
where $\partial_{\alpha}$ indicates the derivative with respect to the parameter $\theta_{\alpha}$ and $f_\mathrm{sky}$ is the fraction of the sky covered by (the intersection of) both surveys. Notice that, since we are not interested in the cosmological parameters but only on the GW bias, we marginalise over all other parameters, also using a prior for the six standard cosmological parameters coming from Planck\footnote{\href{http://pla.esac.esa.int/pla/}{http://pla.esac.esa.int/pla/}} data \cite{Planck:XIII}. 
In this first approach the significance for distinguishing a fiducial model (hereafter ``fiducial'') from an alternative one (hereafter ``alternative'') is given by the  difference between the effective GWs bias parameters (see section \ref{sec:tracers}), after marginalising over all other parameters:
\begin{equation}
\left(\frac{S}{N}\right)^2_{\Delta B/B} = \frac{\left(B^\mathrm{Alternative}_\mathrm{GW}-B^\mathrm{Fiducial}_\mathrm{GW}\right)^2}{\sigma^2_{B^\mathrm{Fiducial}_\mathrm{GW}}},
\label{eq:SN_Fisher}
\end{equation}
where $B_\mathrm{GW}$ is the effective bias defined in Equation \eqref{eq:bias_def} and $\sigma_{B^\mathrm{Fiducial}_\mathrm{GW}}$ is the Fisher-estimated  marginal error on $B_\mathrm{GW}$. 

In the second way we quantify the distance of an alternative model from the fiducial using a $\Delta\chi^2$ statistics. In our case the $\Delta\chi^2$ is given by the logarithm of a likelihood, in particular we assume a likelihood quadratic in the angular power spectra. The resulting $\Delta\chi^2$ statistics reads as
\begin{equation}
\left(\frac{S}{N}\right)^2_{\sqrt{\Delta\chi^2}} \sim \Delta\chi^2 := f_{\mathrm{sky}}\sum_2^{\ell_\mathrm{max}} (2\ell+1) (\mathbf{C}^\mathrm{Alternative}_\ell-\mathbf{C}^\mathrm{Fiducial}_\ell)^T \mathrm{Cov}^{-1}_\ell (\mathbf{C}^\mathrm{Alternative}_\ell-\mathbf{C}^\mathrm{Fiducial}_\ell),
\label{eq:delta_chi2}
\end{equation}
where the vector $\mathbf{C}_\ell$ contains the same data of the covariance matrix $\mathcal{C}_\ell$ in equation \eqref{eq:covariance_matrix_first_method} but organized as
\begin{equation}
\mathbf{C}_\ell = 
\left( \begin{matrix}
C_\ell^{\mathrm{g}\mathrm{g}}(z_1,z_1) \\
\vdots	\\
C_\ell^{\mathrm{g}\mathrm{GW}}(z_1,z_1) \\
\vdots	\\
C_\ell^{\mathrm{GW}\mathrm{GW}}(z_1,z_1) \\
\vdots
\end{matrix} \right),
\label{eq:vectors}
\end{equation}
and where the $\mathrm{Cov}_\ell$ is a new covariance matrix, computed from angular power spectra of the fiducial model. We can associate to every element $I=1,...,21$ of the $\mathbf{C}_\ell$ vector two indices $(I_1,I_2)$, corresponding to the two tracers (in a given redshift bin) that produce the angular power spectra that appear in the $I^\mathrm{th}$ row; for instance we associate to $I=1$, corresponding to $C_\ell^{\mathrm{g}\mathrm{g}}(z_1,z_1)$, the couple of indices $(I_1=\mathrm{g}_{z_1}, I_2=\mathrm{g}_{z_1})$. Then the elements of the new covariance matrix $\mathrm{Cov}_\ell$ read as
\begin{equation}
\left(\mathrm{Cov}_{\ell}\right)_{IJ} = \tilde{C}^{I_1J_1}_\ell\tilde{C}^{I_2J_2}_\ell+\tilde{C}^{I_1J_2}_\ell\tilde{C}^{I_2J_1}_\ell,
\label{eq:covariance_matrix_second_method}
\end{equation}
where the $\tilde{C_\ell}$ are those defined in Equation \eqref{eq:tilde_Cls}.  

Notice that in both cases the ability to distinguish between two scenarios can differ according to which model is the alternative model and which one is the fiducial, since the covariance matrix and thus the errors depend (sometimes strongly) on the fiducial model adopted. The Fisher and $\Delta \chi^2$ methods do not have to give the same results because they are two different approximations. The Fisher approach is based on a quadratic approximation of the log-likelihood, estimating its curvature around the fiducial scenario, while the $\Delta \chi^2$ approach is based directly on the log-likelihood and assumes all other parameters are perfectly known. 


\section{Tracers}
\label{sec:tracers}
In this section we describe the two tracers we consider in this work, galaxies and GWs. For cosmological purposes, each of these tracers is characterised by a source number density per redshift bin and square degree $d^2N_X/dz d\Omega$, bias $b_{X}(z)$, magnification bias $s_{X}(z)$ and evolution bias $f^\mathrm{evo}_X(z)$ parameters, which will be defined in detail below. While some of these quantities are uncertain at the moment, in the following we will attempt to keep track of these uncertainty and how they may affect the final results.

In particular, the two methods presented in section \ref{sec:ggwcorrelation} allow us to assess the effect of the uncertainty on the redshift dependence of the bias, in fact in the Fisher analysis case, starting from the source number density and bias, we associate to a given tracer $X$ an \textit{effective} bias parameter
\begin{equation}
B_X = \frac{{\displaystyle \int_{z_\mathrm{min}}^{z_\mathrm{max}} dz\ b_{X}(z) \frac{d^2N_X}{dz d\Omega}}}{{\displaystyle \int_{z_\mathrm{min}}^{z_\mathrm{max}} dz \frac{d^2N_X}{dz d\Omega}}}
\label{eq:bias_def}
\end{equation}
over the entire redshift range $[z_\mathrm{min},z_\mathrm{max}]$ of the survey, while in the $\Delta\chi^2$ statistics case, we assume that $b_X(z)$ is known. We summarize the values of the effective bias for the different tracers, scenarios  and surveys in table \ref{tab:effective_bias}. In the first case we assume that only the averaged, effective bias is the relevant quantity known well enough to be used as a model parameter, neglecting the information coming from its redshift dependence, while in the second one we are exploiting it to maximize the differences between the two models.

\begin{table}
\centering
\begin{tabular}{|c|c|c|c|c|}
\hline
$B_X$	&	Generic	&	EMU	&	DESI	&	SKA	\\
\hline
\hline
Galaxies	&	$1.55$	&	$0.83$	&	$1.37$	&	$1.58$	\\	
\hline
\hline
Primordial - Early Binaries	&	$1.00$	&	$1.00$	&	$1.00$	&	$1.00$	\\
\hline
Primordial - Late Binaries	&	$0.50$	&	$0.50$	&	$0.50$	&	$0.50$	\\
\hline
End-point of Stellar Evolution	&	$1.81$	&	$0.84$	&	$1.53$	&	$1.85$	\\
\hline
\end{tabular}
\caption{Effective bias for different scenarios and surveys calculated according to equation~\eqref{eq:bias_def}.}
\label{tab:effective_bias}
\end{table}

While the procedure to obtain such quantities for galaxies is well established, in the case of GWs this is quite a new field, and an accurate modelling would involve not only the knowledge of the physics behind the merging process but also an understanding of GWs detection efficiency for mergers events detectors. For this initial investigation we have to make assumptions that may need to be revised and improved in the future. For this reason we present a step-by-step introduction in the GWs section \ref{subsec:gw}, guiding the reader through all the details.


\subsection{Galaxies}
\label{subsec:galaxies}
Depending on the experimental set up  under consideration, we choose as luminous tracers   emission-line galaxies in the redshift range $[0.6-1.7]$, targeted by DESI, or star-forming galaxies, targeted by EMU and SKA in the redshift range $[0.0-5.0]$. In particular the latter will be mapped up to relatively high redshift by forthcoming radio surveys, as extensively discussed in Ref. \cite{jarvis:sfgnumberdensity}, where they dominate the total number of sources. For the generic case we use radio galaxies, as in the SKA case.

\begin{figure}
\centerline{
\subfloat{
	\includegraphics[width=1.0\linewidth]{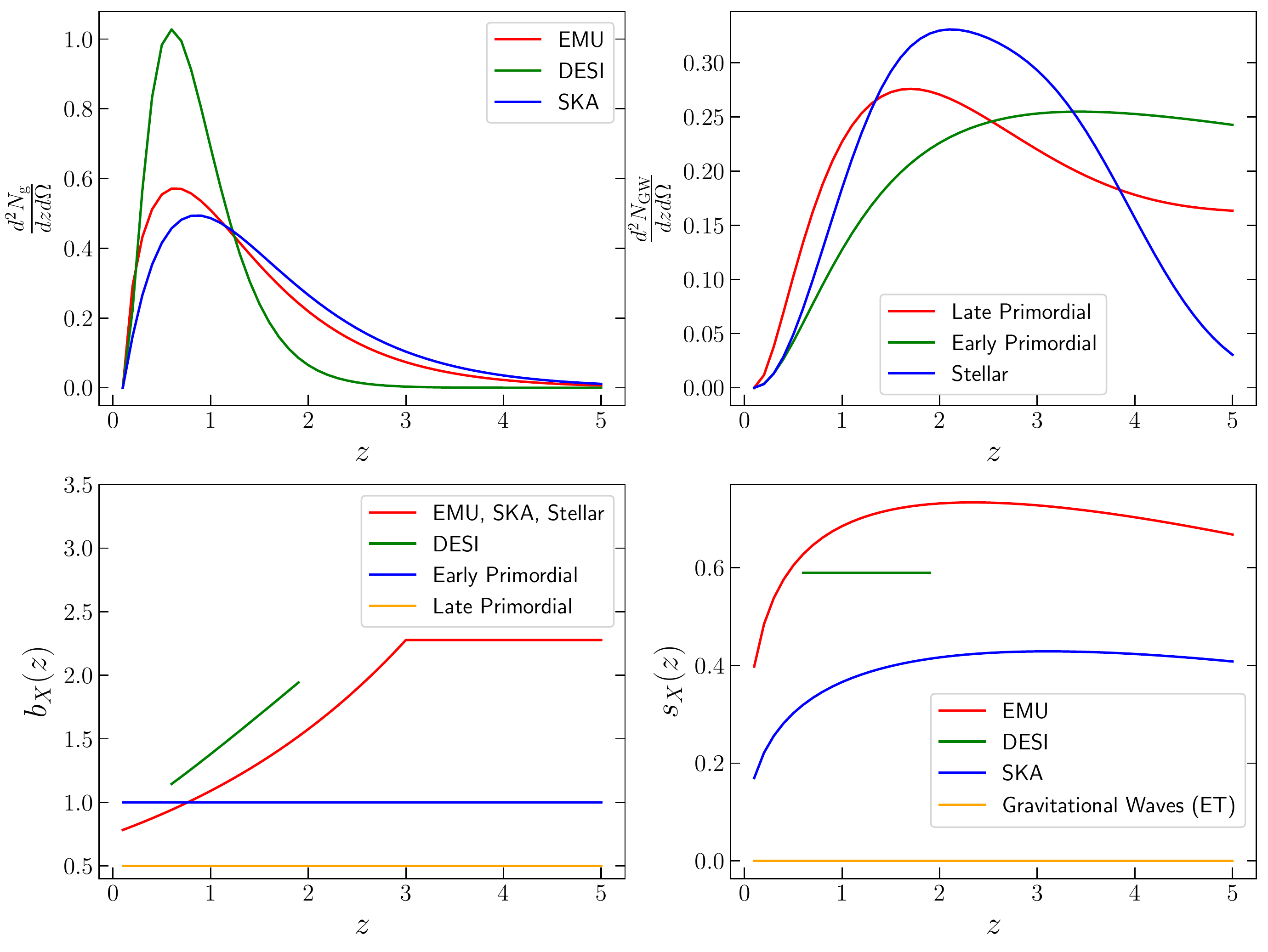}}}
\caption{\textit{Top panels}: normalized number density distribution per redshift bin per square degree $\displaystyle \frac{d^2N_X}{dzd\Omega}$ for galaxies (\textit{top left}) and GWs (\textit{top right}). \textit{Bottom panels}: bias $b_X(z)$ (\textit{bottom left}) and magnification  bias parameter $s_X(z)$ (\textit{bottom right}) for galaxies and GWs. We report the GWs magnification bias  parameter associated to a BHs population with monochromatic mass distribution detected by an interferometer with characteristics similar to those of ET.}
\label{fig:dNdz_bias}
\end{figure}

In all the three surveys we  find that we can model the number density per redshift bin and square degree as
\begin{equation}
\dfrac{d^2N_{\mathrm{g}}}{dz d\Omega} = a_1  z^{a_2}  e^{-a_3z},
\label{eq:dNdz_sfg}
\end{equation}
where different  surveys have different  parameters $a_1,\ a_2,\ a_3$\footnote{ For EMU we find $a_1^\mathrm{EMU}=1236.0$, $a_2^\mathrm{EMU}=0.77$, $a_3^\mathrm{EMU}=1.39$; for DESI we have $a_1^\mathrm{DESI}=56491.0$, $a_2^\mathrm{DESI}=1.89$, $a_3^\mathrm{DESI}=3.70$ while for SKA we have $a_1^\mathrm{SKA}=57642.0$, $a_2^\mathrm{SKA}=1.05$, $a_3^\mathrm{SKA}=1.36$.}. For DESI we used  figure 3.12 of Ref. \cite{Aghamousa:desi}, while for EMU and SKA we used the Tiered Radio Extragalactic Continuum Simulation (T-RECS) \cite{Bonaldi:TRECS} catalogue with different detection threshold ($100\ \mu Jy$ for EMU and $5\ \mu Jy$ for SKA). We report in the top left panel of figure \ref{fig:dNdz_bias} the three normalized number densities. 

The bias for emission-line galaxies is taken to be $b_\mathrm{g}(z)=0.84/D(z)$ \cite{Aghamousa:desi}, where $D(z)$ is the linear growth factor normalized to unity today, while the bias for EMU and SKA star-forming galaxies is modelled as \cite{raccanelli:sfgbias}
\begin{equation}
b_{\mathrm{g}}(z) = 
\left\lbrace \begin{aligned}
& b_0 e^{zb_1}, \ & z < 3,	\\
& b_0 e^{3b_1},  \  & z \geq 3,
\end{aligned} \right.
\label{eq:bias_sfg}
\end{equation}
where $b_0 = 0.755$ and $b_1=0.368$. Following the prescription of Ref. \cite{wilman:radiosourcessimulation}, the bias is assumed to be constant after redshift $z=3$ to avoid unrealistically high values. We show the bias redshift dependence in the bottom left panel of figure \ref{fig:dNdz_bias}. The effective bias of equation \ref{eq:bias_def} for these surveys yields $B_\mathrm{g, EMU}=0.83$, $B_\mathrm{g, DESI}=1.37$ and $B_\mathrm{g, SKA}=1.58$.

Gravitational lensing changes the sources surface density on the sky in two competing ways \cite{turner:magnificationbias}, by increasing the area, which in turn decreases the projected number density, but also by magnifying individual sources and promoting faint objects above the magnitude limit. These effects change the observed number density $n_\mathrm{obs}$ in a flux-limited survey as
\begin{equation}
n_\mathrm{obs} = n_\mathrm{g}\left[ 1 + (5s_\mathrm{g}-2)\kappa \right],
\end{equation}
where $n_\mathrm{g}$ is the intrinsic galaxies number density, $s_\mathrm{g}$ is called galaxy magnification bias parameter and $\kappa$ is the convergence \cite{bartelmann:convergence}, namely an isotropic increase or decrease of source size, defined as $\kappa=\frac{1}{2}\nabla^2\psi$, where $\psi$ is the lensing potential. The change in the number of observed sources depends the value of the slope of the faint-end of the luminosity function \cite{hui:magnificationbias, liu:magnificationbias, montanari:magnificationbias}
\begin{equation}
s_\mathrm{g}(z) = \left.\frac{d\log_{10}\frac{d^2N_\mathrm{g}(z,m<m_\mathrm{lim})}{dzd\Omega}}{dm}\right|_{m_\mathrm{lim}} = -\frac{2}{5}\left.\frac{d\log_{10}\frac{d^2N_\mathrm{g}(z,L>L_\mathrm{lim})}{dzd\Omega}}{d\log_{10}L}\right|_{L_\mathrm{lim}},
\label{eq:galaxymagnificationbias}
\end{equation}
where $m$ is the apparent magnitude, $L$ is the intrinsic source luminosity and $m_\mathrm{lim},\ L_\mathrm{lim}$ are the maximum detectable magnitude and the minimum detectable source luminosity of the survey. The magnification bias enters in the velocity, lensing and gravity terms of equation \eqref{eq:numbercount_fluctuation} (see also appendix \ref{app:relativistic_number_counts}), however its main contribution is in the lensing part, which dominates the amplitude of the signal in the cross-bin case. The reader should keep in mind that the specific value $s_\mathrm{g}=0.4$ is the one associated to a compensation between the two competing effects, therefore it is the one that cancels lensing contributions. For DESI we use figure 3.11 of Ref.~\cite{Aghamousa:desi}, while for EMU and SKA we use the T-RECS catalogue \cite{Bonaldi:TRECS} to compute it. We report the magnification bias parameter $s_\mathrm{g}(z)$ in the bottom right panel of figure \ref{fig:dNdz_bias}.

Finally, it should be noted that the number of galaxies does not have to be conserved as function of redshift, e.g., galaxies can form, therefore their number density does not scale as $a^{-3}$, where $a$ is the scale factor. To account for the creation of new galaxies we include also the evolution bias $f^\mathrm{evo}_X$ defined as \cite{challinor:evolutionbias, jeong:evolutionbias, bertacca:evolutionbias}
\begin{equation}
f^\mathrm{evo}_\mathrm{g}(z) = \frac{d\log\left(a^3\frac{d^2N_\mathrm{g}}{dzd\Omega}\right)}{d\log a}.
\label{eq:evolution_bias}
\end{equation}
This term enters in the velocity and gravity contributions in equation \eqref{eq:numbercount_fluctuation} (see also appendix~\ref{app:relativistic_number_counts}). Since it appears only in subleading terms and since there are significant uncertainties in the modelling of galactic evolution, we can use in the definition of evolution bias the observed number density instead of the true one, without adding significant errors.


\subsection{Gravitational Waves}
\label{subsec:gw}
The number density of detected GWs events per redshift bin per square degree can be estimated as
\begin{equation}
\frac{d^2N_{\mathrm{GW}}}{dz d\Omega} = T_\mathrm{obs} \frac{c\chi^2(z)}{(1+z)H(z)}\mathcal{R}_\mathrm{tot}(z)F^\mathrm{detectable}_{\mathrm{GW}}(z),
\label{eq:dNdz_gw}
\end{equation}
where $T_\mathrm{obs}$ is the total observational time\footnote{We assume as fiducial observational time $T_\mathrm{obs}=10$ years.}, $\chi(z)$ is the comoving distance, $H(z)$ is the Hubble expansion rate, $\mathcal{R}_\mathrm{tot}(z)$ is the total comoving merger rate, $F^\mathrm{detectable}_{\mathrm{GW}}(z)$ is the fraction of detectable events, that depends on the Signal-to-Noise cut $\varrho_\mathrm{lim}$ imposed at the GWs observatory. The total merger rate depends on the progenitors origin. In section \ref{subsubsec:primordial_case} we consider a scenario where mergers are from PBHs; in section \ref{subsubsec:stellar_case} a scenario with BHs of stellar origin. We comment in section \ref{subsec:gw_SNR} how we compute $F_\mathrm{GW}^\mathrm{detectable}(z)$. Starting from the observed number density, we can calculate GWs evolution bias $f^\mathrm{evo}_\mathrm{GW}(z)$ using equation \eqref{eq:evolution_bias}.
 
Here we should mention that the uncertainty in the total merger rate $\mathcal{R}_\mathrm{tot}$ is of orders of magnitude, however what enters in the calculation of the angular power spectra $C^{XY}_\ell$ (i.e., the signal) is the shape of $\frac{d^2N}{dzd\Omega}$, not the global amplitude. On the other hand, the merger rate (and its normalisation) affects the signal-to-noise ratio (i.e., the error-bars): a larger number density will decrease the shot noise, improving the constraints on the cosmological parameters of interest.


\subsubsection{Primordial Scenario}
\label{subsubsec:primordial_case}
In the case where BHs have primordial origin and form a significant part of the dark matter there are at least two important processes that lead to BHs binary formation. In the \textit{late primordial} formation scenario the binary forms when progenitors are already part of dark matter halos \cite{bird:pbhasdarkmatter} and become a bound system by emitting GWs, while in the \textit{early primordial} formation scenario the bound pair forms during radiation-dominated era \cite{nakamura:pbhmergerrate, alihaimoud:pbhmergerrate, raidal:earlybinary}. In particular, the first mechanism, effective at late times, yields merger rate compatible with those of LIGO. The second mechanism, according to theoretical estimates done in Ref. \cite{alihaimoud:pbhmergerrate}, provides a preferred present-day merger rate thait is already excluded by the LIGO constraints.  Nevertheless, given the large theoretical uncertainties that these estimates involve, we keep both scenarios into account, adjusting the merger rate value of the early primordial case to make it compatible with LIGO current constraints (we discuss this in more details in section \ref{sec:results}).  

In the following we focus on the late primordial scenario and we briefly review the theoretical modelling of Ref. \cite{bird:pbhasdarkmatter}, reporting only the most important results and extending their formalism to the case where PBHs have an extended mass distribution instead of a monochromatic one. In this model, the total merger rate $\mathcal{R}_\mathrm{tot}(z)$ is expressed as a function of the merger rate per halo $\mathcal{R}_{\mathrm{halo}}$ as
\begin{equation}
\mathcal{R}_\mathrm{tot}(z) = \int_{M_\mathrm{halo,min}}^{M_\mathrm{halo,max}} dM_\mathrm{halo} \frac{dn(M_\mathrm{halo},z)}{dM_\mathrm{halo}}\mathcal{R}_{\mathrm{halo}}(M_\mathrm{halo},z),
\label{eq:totalmergerrate}
\end{equation}
where $M_\mathrm{halo,min},M_\mathrm{halo,max}$ are the minimum and maximum mass of dark matter halos and $\displaystyle \frac{dn(M_\mathrm{halo},z)}{dM_\mathrm{halo}}$ is the halo mass function \cite{tinker:halomassfunction}. Following the notation of Ref. \cite{bellomo:emdconstraints} and extending their formalism, for a completely general PBHs mass distribution, the merger rate per halo is given by
\begin{equation}
\mathcal{R}_{\mathrm{halo}}(M_\mathrm{halo},z) = f^2_\mathrm{PBH} \int d^3r dM_\mathrm{1} dM_\mathrm{2} \frac{d\Phi_\mathrm{PBH}}{dM_\mathrm{1}} \frac{d\Phi_\mathrm{PBH}}{dM_\mathrm{2}} \frac{\left\langle \sigma_\mathrm{PF} v_\mathrm{PBH}\right\rangle}{2M_\mathrm{1}M_\mathrm{2}} \rho_\mathrm{halo}^2(r),
\label{eq:mergerrateperhalo}
\end{equation}
where $f_\mathrm{PBH}$ is the fraction of dark matter composed by PBHs\footnote{In this work we assume that PBHs compose the totality of dark matter, i.e., $f_\mathrm{PBH}=1$.}, $\frac{d\Phi_\mathrm{PBH}}{dM}$ describes the shape of the mass distribution and is normalized to unity, $M_1,M_2$ are progenitors masses, $\sigma_\mathrm{PF}$ is the  pair-formation cross section \cite{mouri:pbhpairformation, quinlan:pbhpairformation}, $v_\mathrm{PBH}$ is the relative velocity between two PBHs, angle brackets $\left\langle\ \cdot\ \right\rangle$ stand for the average over PBHs relative velocity distribution (a Maxwell-Boltzmann distribution with a cut-off at virial velocity) and $\rho_\mathrm{halo}(r)$ is the halo radial profile which we choose to be a Navarro-Frenk-White (NFW) \cite{navarro:nfwhaloprofile}. The NFW profile  is governed by the so-called concentration parameter that can be calibrated on numerical simulations \cite{prada:concentrationparameter, ludlow:concentrationparameter}, in turn the ``typical'' value of the concentration parameter depends only the halo mass and on the redshift. Since the pair-formation cross section $\sigma_\mathrm{PF}$ scales as $(M_1+M_2)^{10/7}M_1^{2/7}M_2^{2/7}$ with progenitors masses, the black holes mass dependence of the halo merger rate can be factorized as
\begin{equation}
\mathcal{M}_\mathrm{PBH} = \int dM_1dM_2\frac{d\Phi_\mathrm{PBH}}{dM_1}\frac{d\Phi_\mathrm{PBH}}{dM_2}\frac{(M_1+M_2)^{10/7}}{M_1^{5/7}M_2^{5/7}}.
\label{eq:emdmergerrate}
\end{equation}
Therefore, in the case where the two merging objects come from a monochromatic mass distribution, i.e., a Dirac delta centred at a certain value $M_\mathrm{PBH}$, equation \eqref{eq:emdmergerrate} simplifies to $\mathcal{M}_\mathrm{PBH}^\mathrm{Monochromatic} = 4^{5/7}$, independently from where the mass distribution is centred.

\begin{figure}[h!]
\centerline{
\subfloat{
	\includegraphics[width=0.5\linewidth]{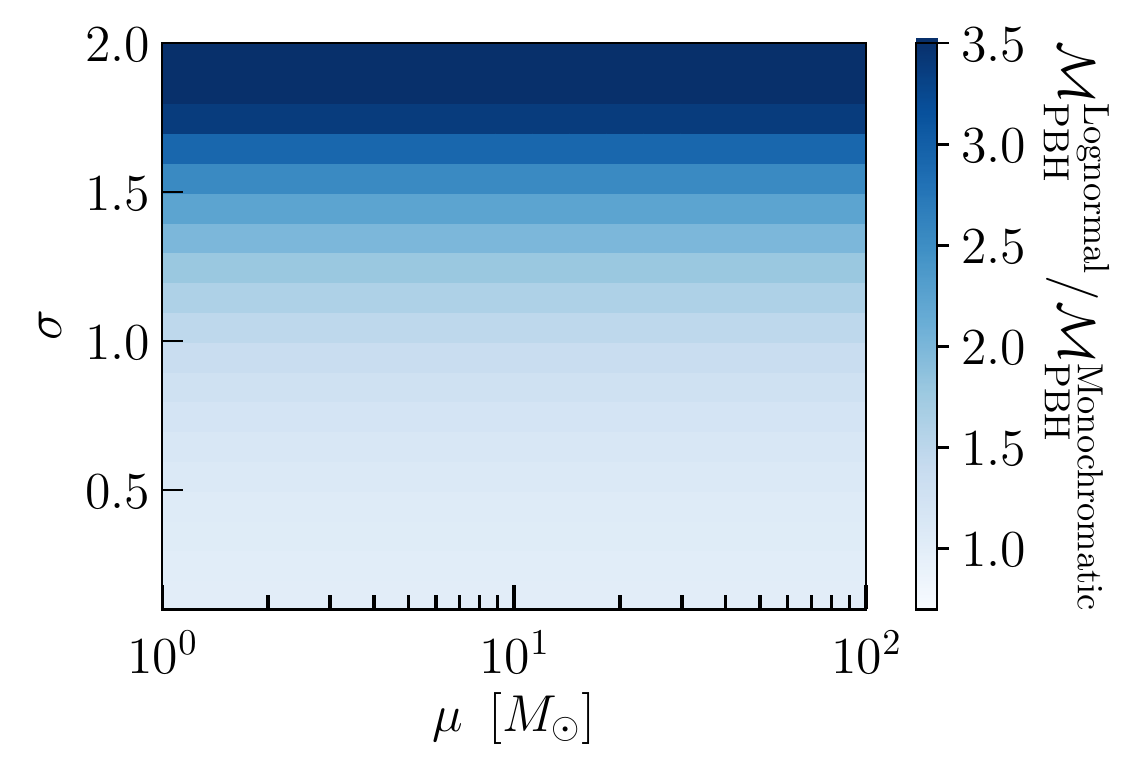}}}
\centerline{	
\subfloat{
	\includegraphics[width=1.0\linewidth]{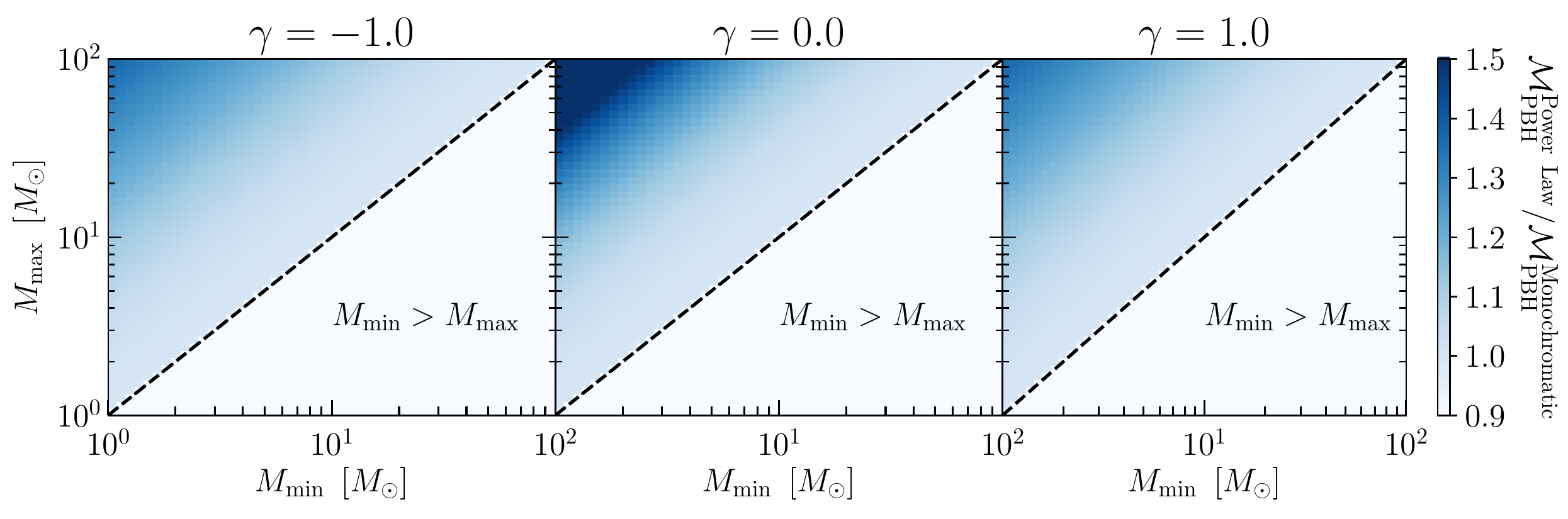}}}
\caption{Ratio between the mass-dependent factor $\mathcal{M}_\mathrm{PBH}$ of different extended mass distribution, centred in the $\mathcal{O}(10)\ M_\odot$ window, with respect to the monochromatic case $\mathcal{M}^\mathrm{Monochromatic}_\mathrm{PBH}=4^{5/7}$. \textit{Top panel:} \textit{Lognormal} distribution. \textit{Bottom panel:} \textit{Power Law} distribution. In both cases the $\mathcal{M}_\mathrm{PBH}$ factor does not deviate significantly from the one obtained in the monochromatic case.}
\label{fig:emdmergerrate}
\end{figure}

Since perfectly monochromatic mass distributions are unphysical, we estimate how the merger rate in this model changes when considering two popular extended mass distributions (see Ref. \cite{bellomo:emdconstraints} for more details about them). We focus on \textit{Power Law} distributions
\begin{equation}
\frac{d\Phi_\mathrm{PBH}}{dM} = \frac{\mathcal{N}_{PL}}{M^{1-\gamma}}\Theta(M-M_\mathrm{min})\Theta(M_\mathrm{max}-M),
\end{equation}
characterized by an exponent $\gamma\in [-1,+1]$, a mass range $(M_\mathrm{min}, M_\mathrm{max})$ and a normalization factor $\mathcal{N}_{PL}$ and on \textit{Lognormal} distributions
\begin{equation}
\frac{d\Phi_\mathrm{PBH}}{dM} = \frac{e^{-\frac{\log^2(M/\mu)}{2\sigma^2}}}{\sqrt{2\pi}\sigma M},
\end{equation}
where $\log\mu$ and $\sigma$ are the mean and standard deviation of the logarithm of the mass, respectively. It is not possible to find an analytical result of the integral in equation \eqref{eq:emdmergerrate} for these two distributions, however we provide in figure \ref{fig:emdmergerrate} the ratio between merger rates calculated for an extended and a monochromatic mass distribution, which is equivalent to the ratio $\mathcal{M}_\mathrm{PBH}^\mathrm{Extended}/\mathcal{M}_\mathrm{PBH}^\mathrm{Monochromatic}$ between the factors calculated in equation \eqref{eq:emdmergerrate}. As can be appreciated from the figure, the ratio is always a factor few bigger than unity. This important result allows us to generalize to the extended mass distribution case (for distributions centred in the $\mathcal{O}(10)\ M_\odot$ window) conclusions we draw for the monochromatic one in section \ref{sec:results}, just by rescaling the merger rate by some numerical factor. Furthermore, in the case of a \textit{Lognormal} distribution, the ratio depends on the width of the distribution but not on the ``scale'' $\mu$.\footnote{This result is exact, in fact the scale $\mu$ disappears from equation \eqref{eq:emdmergerrate} once that we rescale the masses through a change of variables.}

Since the total merger rate in equation \eqref{eq:totalmergerrate} is dominated by low-mass halo, the choice of the minimum halo mass may have a big impact on the final result. The minimum mass is set by requiring that the small halos evaporation time is larger the Hubble time. The authors of Ref. \cite{alihaimoud:pbhmergerrate} pointed out  that the presence of Poisson fluctuations significantly affects the estimation of the characteristic density and velocity dispersion, increasing the initial estimate of the minimum allowed halo mass in Ref. \cite{bird:pbhasdarkmatter} by one order of magnitude. However, still in Ref. \cite{alihaimoud:pbhmergerrate}, it was also found that the total merger rate computed including Poisson fluctuations is of the same order of magnitude of that found in Ref. \cite{bird:pbhasdarkmatter}. These results have been obtained assuming that the initial clustering of PBHs is Poissonian, however this assumption is still a matter of discussion. If PBHs are born strongly clustered, merger rate estimates could be heavily affected, see e.g., Refs. \cite{chisholm:pbhclusteringI, chisholm:pbhclusteringII, alihaimoud:pbhclustering, desjacques:pbhclustering, ballesteros:pbhclustering}.

The second scenario we consider is the early primordial formation mechanism. In this case PBHs binaries form in the early Universe and can merge in less than a Hubble time. The estimated total merger rate for this scenario is five orders of magnitude larger than the one given by the late primordial scenario, however there are several theoretical uncertainties; the interested reader can check Ref. \cite{alihaimoud:pbhmergerrate} to find a broad discussion on these uncertainties. Finally we mention that other binary formation mechanisms exist, see e.g., Ref. \cite{nishikawa:pbhmergers}, however we do not consider them since uncertainties at play are even larger.
     
Let's turn our attention on the bias the GWs associated to different binary formation mechanism. If the progenitors have primordial origin, the estimates in Ref. \cite{alihaimoud:pbhmergerrate} indicate that the merger rate is heavily dominated the early primordial one. In this scenario the PBHs trace the dark matter, therefore they have constant bias $b_\mathrm{GW}(z)\equiv 1$ and constant effective bias $B_\mathrm{GW}\equiv 1$. However,  if primordial binaries are disrupted during the history of the Universe and the merger rate is dominated by the late primordial mechanism, then the merger events trace low-velocity dispersion low-mass halos ($M_\mathrm{halo}<10^6M_\odot$). The bias of these halos is given by \cite{mo:smallhalosbias}
\begin{equation}
b_\mathrm{lmh}(z) = 1 + \dfrac{\nu^2-1}{\delta_\mathrm{sc}},
\label{eq:bias_GW}
\end{equation}
where $\nu=\delta_\mathrm{sc}/\sigma(M_\mathrm{halo},z)$ is the dimensionless peak height, $\delta_\mathrm{sc}=1.686$ is the spherical collapse threshold and $\sigma(M_\mathrm{halo},z)$ is the root-mean-squared density fluctuation at redshift z for a smoothing scale corresponding to  mass $M_\mathrm{halo}$ in a Press-Schechter-like philosophy. Since the bias does not evolve significantly in redshift, we consider it as a constant, therefore in this second case its effective value is $B_\mathrm{GW}=0.5$.


\subsubsection{End-point of Stellar Evolution Scenario}
\label{subsubsec:stellar_case}
Several authors have estimated the merger rate of BH binaries coming from stellar evolution, see e.g., Refs. \cite{belczynsky:mergerrate, elbert:bhmergerrate, Mapelli:merger_rate}. In this work we use the prescriptions given by Ref. \cite{Mapelli:merger_rate} who obtain the merger rate of stellar BH binaries by combining the Illustris cosmological simulation with population-synthesis simulations of black hole binaries. They rely on up-to-date prescriptions for stellar winds and core collapse Supernovae. In this merger rate all three populations of stars are included. We consider as our fiducial model the fiducial merger rate of figure 1 of \cite{Mapelli:merger_rate}. We can model their merger rate as
\begin{equation}
\mathcal{R}_\mathrm{tot}(z) = \mathcal{A}\left(1+\left(\frac{z}{z_0}\right)^{p_1}\right)^{p_2} e^{-(z-z_1)^2/2}
\end{equation}
where $\mathcal{A}=786.0$, $z_0=3.0$, $z_1=1.8$, $p_1 = 4.9$ and $p_2 = 1.4$. Other models (based on different properties of the population-synthesis simulations, using different prescriptions for Supernovae, natal kicks distribution, Hertzsprung gap stars and common envelope phase efficiency) still have approximately the same shape, but a different amplitude $\mathcal{A}$. Uncertainties in this case are around one order of magnitude. We report the normalized number density of sources per redshift bin per square degree compute using this total merger rate in the top right panel of figure \ref{fig:dNdz_bias}.

When progenitors of a merging event have stellar origin, they are more likely correlated with higher-mass halos that had a higher star-formation rate, therefore their bias will be the same of the  galaxies under consideration, i.e. $b_{\mathrm{GW}}(z) = b_{\mathrm{g}}(z)$, where the different bias are reported in section \ref{subsec:galaxies} and figure \ref{fig:dNdz_bias}. In these cases the effective bias reads as $B_\mathrm{GW,EMU}=0.84$, $B_\mathrm{GW,DESI}=1.53$ and $B_\mathrm{GW,SKA}=1.85$.

  
\subsubsection{Gravitational Waves Signal-to-Noise Ratio and Event Detectability}
\label{subsec:gw_SNR}
In this section we calculate the expected signal-to-noise ratio for BH-BH merger events. Given the uncertainties in the final design of future GWs observatories, we take several simplifying assumptions. However, we try to be as realistic as possible striking a balance between being conservative but not over-conservative. Our main findings are robust against changes of specific details.

We define the GW averaged\footnote{Here the average is over the system-detector relative orientation and over waves polarization.} signal-to-noise ratio measured at a given GWs observatory as~$\sqrt{\left\langle\varrho^2\right\rangle}$. This is obtained via (see e.g., Ref. \cite{moore:gwsignaltonoise}),
\begin{equation}
\left\langle\varrho^2\right\rangle = \frac{1}{5}\int_{f_\mathrm{min}}^{f_\mathrm{max}} df \frac{h_c^2(f)}{f^2 S_n(f)},
\label{eq:SNR_average}
\end{equation}
where $f$ is the observed frequency, $h_c(f)$ is the characteristic strain amplitude, $S_n(f)$ is the one-sided noise power spectral density and $f_\mathrm{min}$ and $f_\mathrm{max}$ are the minimum and maximum frequencies the detector is sensible to. We choose $(f_\mathrm{min},f_\mathrm{max})=(10,10^4)$ when considering aLIGO and  $(f_\mathrm{min},f_\mathrm{max})=(1,10^4)$ when considering ET. To be more precise, the $1/5$ factor slightly depends on the characteristics of the analysed system considered, however it is usually very close to the value we have chosen \cite{vallisneri:gwsnr}.

The characteristic strain amplitude depends on the physical phenomenon one is interested in. In this work we consider a two body system, the merging binary, with progenitors masses $M_1$ and $M_2$, total mass $M_\mathrm{tot}$ and reduced mass $\mu_\mathrm{r}$. The characteristic strain is related to its spectral energy distribution $dE/df_s$ as \cite{flanagan:gwsignaltonoise}
\begin{equation}
h_c(f) = \frac{2^{1/2}}{\pi \chi(z)}\sqrt{\frac{dE}{df_s}},
\end{equation}
where $f_s=(1+z)f$ is the frequency of the emitted wave at the source, located at redshift $z$. Every merger event is characterized by three different phases, inspiraling ($I$), merging ($M$) and ringdown ($R$), that correspond to the emission in three different and separate frequency ranges with different spectral energy distributions (see Ref. \cite{flanagan:gwsignaltonoise} for specific details). In the following we consider only the first two phases (inspiraling and merging) since during the third one the strain is rapidly damped. These phases are associated to the observed frequency ranges
\begin{equation}
I:\quad f<\frac{f_m}{1+z},\qquad M:\quad \frac{f_m}{1+z}<f<\frac{f_r}{1+z},
\end{equation}
where $f_m=4100\ (M_\odot/M_\mathrm{tot})\ s^{-1}$ and $f_r=28600\ (M_\odot/M_\mathrm{tot})\ s^{-1}$ are the so called merger and ringdown frequencies. 

Given that different merger phases do not share the same frequency domain, the signal-to-noise ratio can be computed separately for each phase  and then combined:
\begin{equation}
\begin{aligned}
\left\langle\varrho^2\right\rangle &= \left\langle\varrho^2\right\rangle_I  + \left\langle\varrho^2\right\rangle_M, \\
\left\langle\varrho^2\right\rangle_I &= \frac{2\mu_\mathrm{r} M_\mathrm{tot}^{2/3}G^{5/3}}{15\pi^{4/3}c^3}\frac{(1+z)^{-1/3}}{\chi^2(z)}\int_{f_\mathrm{min}}^{f^I_\mathrm{up}(z)} df\frac{f^{-7/3}}{S_n(f)},	\\
\left\langle\varrho^2\right\rangle_M &= \frac{32G\mu^2_\mathrm{r}\epsilon}{5\pi^2M_\mathrm{tot}c(f_r-f_m)}\frac{1}{\chi^2(z)}\int_{f^M_\mathrm{low}(z)}^{f^M_\mathrm{up}(z)} df\frac{f^{-2}}{S_n(f)},
\end{aligned}
\label{eq:snr}
\end{equation}
where $\epsilon$ is the fraction of the total mass emitted in GW and typically assumes values around $\epsilon\sim 0.05$ (this is the value we assume). The signal-to-noise ratio for the different phases are non-zero only if $f_\mathrm{min}<f^I_\mathrm{up}(z)$ and $f^M_\mathrm{low}(z)<f^M_\mathrm{up}(z)$, where
\begin{equation}
f^I_\mathrm{up}(z)=\mbox{min}\left(f_\mathrm{max},\frac{f_m}{1+z}\right), \quad f^M_\mathrm{low}(z)=\mbox{max}\left(f_\mathrm{min},\frac{f_m}{1+z}\right), \quad f^M_\mathrm{up}(z)=\mbox{min}\left(f_\mathrm{max},\frac{f_r}{1+z}\right).
\end{equation}
Each  of the observatories we consider (aLIGO starting in the next decade and, beyond that, ET)  is characterised by its sensitivity or, equivalently, by its own noise power spectral density $S_n(f)$. The interested reader can find these details in Ref. \cite{ajith:noisepsd} for aLIGO and in Ref. \cite{huerta:etnoise} for ET.

To define the fraction of detectable events $F_\mathrm{GW}^\mathrm{detectable}$ introduced in equation \eqref{eq:dNdz_gw}, we impose $\sqrt{\left\langle\varrho^2\right\rangle}>\varrho_\mathrm{lim}$; following existing literature we choose the typical minimum value $\varrho_\mathrm{lim}=8$. The distribution of detected signal-to-noise ratio at different redshift, that we denote with $\mathcal{F}\left(\sqrt{\left\langle\varrho^2\right\rangle}, z\right)$, expected from different BHs mass distributions is computed as follows. 

In the case where the BHs have a monochromatic mass distribution, all the mergers have the same averaged signal-to-noise ratio at a given redshift, therefore in this approximation we are able to detect all the merger events up to some maximum redshift $z_\mathrm{max}$ such that $\sqrt{\left\langle\varrho^2(z_\mathrm{max})\right\rangle}=\varrho_\mathrm{lim}$. We calculate that $z^\mathrm{aLIGO}_\mathrm{max}=0.4$ for aLIGO and $z^\mathrm{ET}_\mathrm{max}>5$ for ET. In this case $F_{\rm GW}^{\rm detectable}\equiv 1$ up to the maximum redshift.

\begin{figure}[h!]
\centerline{
\subfloat{
	\includegraphics[width=0.49\linewidth]{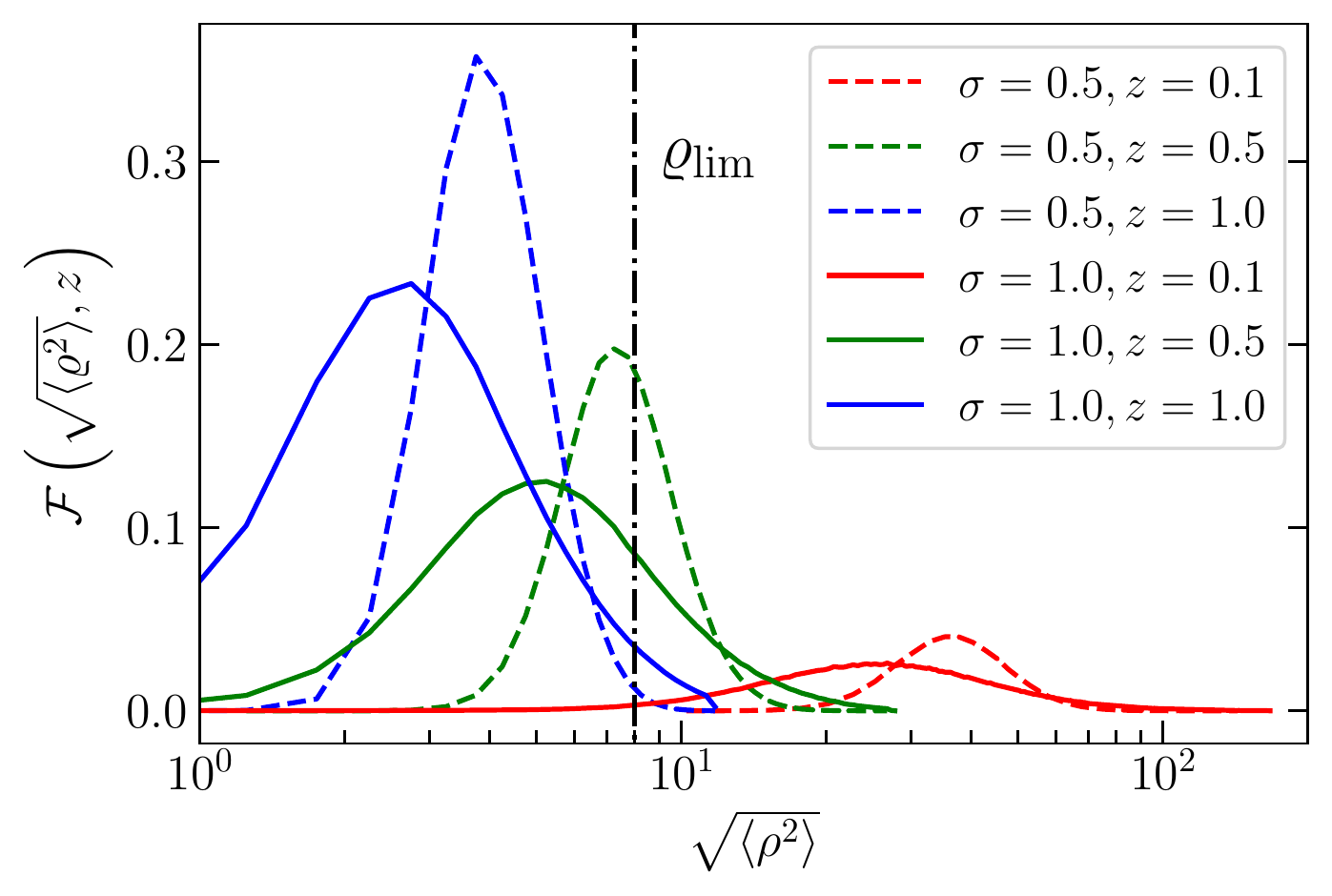}}
\subfloat{
	\includegraphics[width=0.50\linewidth]{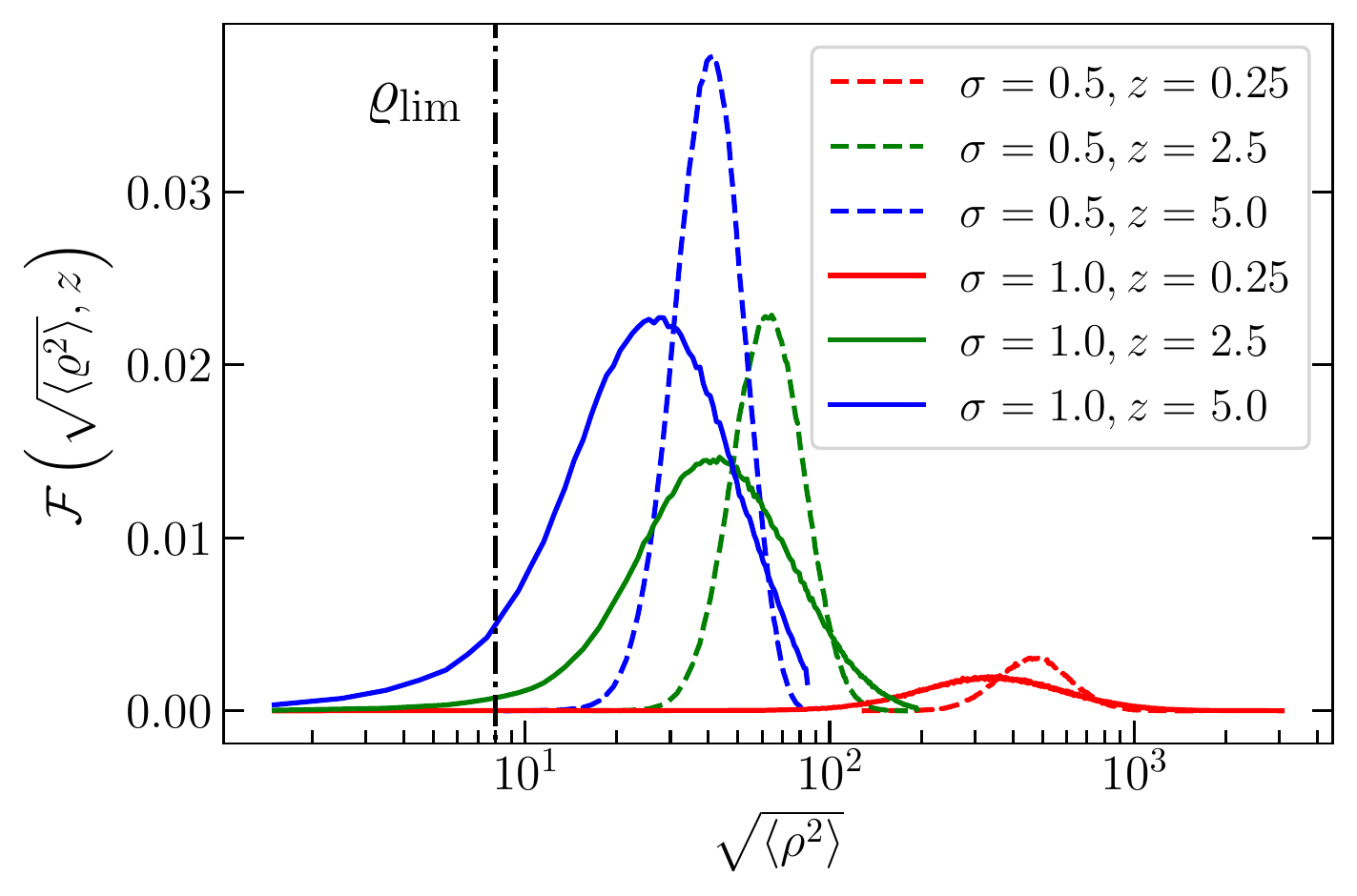}}}
\caption{Probability distribution function of the signal-to-noise for Lognormal black holes extended mass distribution with $(\mu,\sigma)=(30,0.5)$ (\textit{dashed line}) and $(\mu,\sigma)=(30,1.0)$ (\textit{solid line}) at different redshift. \textit{Left panel:} aLIGO detector. \textit{Right panel:} ET detector. The black dotted-dashed line represents the detection threshold at $\varrho_\mathrm{lim}=8$.}
\label{fig:noisedistribution}
\end{figure}

For an extended mass distribution we simulate $10^5$ BH mergers at different redshift with progenitors masses drawn from two \textit{Lognormal} mass distributions, one narrow ($\sigma=0.5$), the other wider ($\sigma=1.0$), both having $\mu=30.0\ M_\odot$. We calculate the averaged signal-to-noise ratio distribution using the aLIGO \cite{ajith:noisepsd} and ET \cite{huerta:etnoise} expected sensitivity. We report the probability distribution $\mathcal{F}\left(\sqrt{\left\langle\varrho^2\right\rangle}, z\right)$ in figure \ref{fig:noisedistribution} for different redshift. For the aLIGO detector details of the extended mass distribution can be relevant, in fact in both cases of narrow and broad mass distribution, a significant fraction of events  may lie below detection threshold, up to half of the total events in the broad distribution case for redshift $z=0.4$, corresponding to $F_\mathrm{GW}^\mathrm{detectable}(z=0.4)\simeq 0.5$. However, the next generation of GWs observatories will overcome this limitation. In particular ET sensitivity will be so large that in the narrow mass distribution case all the event are detectable, therefore $F_\mathrm{GW}^\mathrm{detectable}\equiv 1$.  In the broad mass distribution case we observe that  part of the tail is below the threshold, however even at redshift $z=5$ the overall effect is very small, in fact we find that $F_\mathrm{GW}^\mathrm{detectable}\gtrsim 0.97$. Since in the ET case the effect of an extended mass distribution is so small, every conclusion we draw for the monochromatic case applies also the extended one. 

The GW averaged signal-to-noise ratio and the distribution of detected signal-to-noise ratio as a function of redshift are also a key inputs to compute the magnification bias.


\subsubsection{Gravitational Waves Magnification Bias}
\label{subsubsec:gw_magnification_bias}
In this section we calculate for the first time the magnification bias for GWs. As for galaxies, the magnification bias contains the information on which of the two effects of gravitational lensing, explained in section \ref{subsec:galaxies}, dominates. It enters in the velocity, lensing and gravity terms of equation \eqref{eq:numbercount_fluctuation} (see also appendix \ref{app:relativistic_number_counts}).

In complete analogy to what has been done for galaxies in equation \eqref{eq:galaxymagnificationbias}, we identify the galaxy apparent magnitude $m$ with the GWs averaged signal-to-noise ratio $\sqrt{\left\langle\varrho^2\right\rangle}$ measured at a given GWs observatory. As in equation \eqref{eq:galaxymagnificationbias}, instead of a maximum apparent magnitude $m_\mathrm{lim}$, we define a minimum average signal-to-noise ratio $\varrho_\mathrm{lim}=8$ to claim the detection of an event. Finally, we define the GWs magnification bias parameter as
\begin{equation}
\begin{aligned}
s_\mathrm{GW}(z) &= -\left.\frac{d\log_{10}\frac{d^2N_\mathrm{GW}(z,\sqrt{\left\langle\varrho^2\right\rangle}>\varrho_\mathrm{lim})}{dzd\Omega}}{d\sqrt{\left\langle\varrho^2\right\rangle}}\right|_{\varrho_\mathrm{lim}},
\end{aligned}
\label{eq:gwmagnificationbias}
\end{equation}
where the minus sign has been introduced to preserve the interpretation of a positive or negative magnification bias parameter. Notice that instead of the averaged signal-to-noise ratio $\sqrt{\left\langle\varrho^2\right\rangle}$ one could have defined the GWs magnification bias using just the signal-to-noise ratio $\varrho$. In that case one should consider not only the mass distribution but also the system orientation and GWs polarization distribution. This goes beyond the purposes of this article and is left to future work.

The distribution of detected signal-to-noise ratio at different redshifts $\mathcal{F}\left(\sqrt{\left\langle\varrho^2\right\rangle}, z\right)$ and different BHs mass distributions (see section \ref{subsec:gw_SNR}) is the key quantity to estimate the GWs magnification bias. For a  monochromatic mass distribution, all the mergers have the same averaged signal-to-noise ratio at a given redshift and  we observe all the events (or none of them), hence the magnification bias parameter for a BHs monochromatic population is identically zero ($s_\mathrm{GW}(z)\equiv 0.0$) except in an infinitely thin redshift shell around $z_\mathrm{max}$.

On the other hand, for an extended mass distribution, the magnification bias parameter can be non-zero since, especially for the broad mass distribution, in fact we potentially have events above and below the detection threshold at every redshift. As shown in figure~\ref{fig:noisedistribution} and discussed in section~\ref{subsec:gw_SNR}, the ET sensitivity guarantees that for narrow mass distribution all events are detectable, $s_\mathrm{GW}\equiv 0.0$, as in the monochromatic mass distribution case. In the broad mass distribution case only few events are missed and we find $s_\mathrm{GW}\lesssim 0.01$ at redshift $z=5$. For the aLIGO case we find $s_\mathrm{GW}\lesssim 0.07$ at redshift $z=0.4$ in both cases. In general, for any given extended mass distribution, more sensitive experiments, as the ET, have smaller magnification bias parameter. Since the values of the magnification bias parameter are so close to zero, the error coming from working with a monochromatic mass distribution are subdominant with respect to other uncertainties in the modelling. Therefore we can safely extrapolate our results for monochromatic cases to extended mass distribution cases.


\section{Results}
\label{sec:results}
Beside the tracer's bias, the merger rates are also poorly known, both on the observational and theoretical side, spanning several orders of magnitude and affecting the overall expected number of GWs events. After the first run of LIGO, the observational merger rate today is estimated to be \cite{abbott:ligomergerrates}
\begin{equation}
\mathcal{R}^\mathrm{LIGO}_\mathrm{today} \simeq 9-240\ \mathrm{Gpc^{-3}yr^{-1}},
\end{equation}
while the theoretically predicted merger rates for the stellar and late primordial scenario are
\begin{equation}
\begin{aligned}
\mathcal{R}^\mathrm{Stellar}_\mathrm{today} &\simeq 150 \ \mathrm{Gpc^{-3}yr^{-1}},   \\
\mathcal{R}^\mathrm{Late\ Primordial}_\mathrm{today} &\simeq 4 \ \mathrm{Gpc^{-3}yr^{-1}}.
\end{aligned}
\label{eq:theoretical_merger_rates}
\end{equation}
The predicted merger rate for the early primordial scenario is approximately $10^5 \ \mathrm{Gpc^{-3}yr^{-1}}$ or even higher \cite{hayasaki:mergerrate}, therefore ruling out this scenario. However, given the high uncertainties in its computation which could significantly lower this value \cite{alihaimoud:pbhmergerrate}, in this work we consider a fiducial value for this scenario of 
\begin{equation}
\mathcal{R}^\mathrm{Early\ Primordial}_\mathrm{today} \simeq 200 \ \mathrm{Gpc^{-3}yr^{-1}},
\label{eq:theoretical_merger_rates_early}
\end{equation}
which is consistent with LIGO constraints. Nevertheless, these uncertainties act mainly as a rescaling of the number of events, thus of the noise and of the resulting signal-to-noise ratio.

We parametrize the uncertainty on the number of GWs events \eqref{eq:dNdz_gw} by introducing a new parameter $r$ constant in redshift. In the stellar case $r$ reads as
\begin{equation}
r^{\mathrm{Stellar}} = \frac{T_\mathrm{obs}}{10\ \mathrm{years}} \times \frac{\mathcal{R}}{\mathcal{R}_{\mathrm{today}}^{\mathrm{Stellar}}},
\label{eq:r_parameter}
\end{equation}
where $\mathcal{R}/\mathcal{R}_\mathrm{today}^\mathrm{Stellar}$ parametrizes the uncertainty coming from the chosen fiducial model in Ref.~\cite{Mapelli:merger_rate}. In the primordial scenarios we have
\begin{equation}
\begin{aligned}
r^{\mathrm{Late\ Primordial}} &= \frac{T_\mathrm{obs}}{10\ \mathrm{years}} \times f_\mathrm{PBH}^2 \times \frac{\mathcal{M}^\mathrm{Extended}_\mathrm{PBH}}{\mathcal{M}^\mathrm{Monochromatic}_\mathrm{PBH}} \times \left\langle F^\mathrm{detectable}_\mathrm{GW}\right\rangle \times \dfrac{\mathcal{R}}{\mathcal{R}_{\mathrm{today}}^{\mathrm{Late\ Primordial}}},	\\
r^{\mathrm{Early\ Primordial}} &= \frac{T_\mathrm{obs}}{10\ \mathrm{years}} \times \dfrac{\mathcal{R}}{\mathcal{R}_{\mathrm{today}}^{\mathrm{Early\ Primordial}}},
\end{aligned}
\label{eq:r_parameter_pbh}
\end{equation}
where in the late primordial formation scenario we have explicitly separated the contributions analysed in section \ref{subsec:gw}, even if the dependence on the observational time $T_\mathrm{obs}$, the fraction $f_\mathrm{PBH}$ of PBHs that constitutes the dark matter, the choice of PBHs mass distribution (extended or monochromatic) and the average fraction of observables events $\left\langle F^\mathrm{detectable}_\mathrm{GW}\right\rangle$ can be generalized also to the early primordial formation scenario. The quantity $\mathcal{R}/\mathcal{R}_{\mathrm{today}}^{\mathrm{Late,\: Early\ Primordial}}$ contains any possible uncertainty related to the modelling of the merger rate that affects its overall normalisation, expressing deviations from the fiducial values of equations \eqref{eq:theoretical_merger_rates} and \eqref{eq:theoretical_merger_rates_early}. The values $r^{\mathrm{Stellar, \: Late, \: Early\ Primordial}} = 1$ correspond to the merger rates reported in equations \eqref{eq:theoretical_merger_rates} and \eqref{eq:theoretical_merger_rates_early}. To account for several theoretical uncertainties that can influence the merger rates, we provide results for a range $r^{\mathrm{Stellar, \: Late, \: Early\ Primordial}} \in [10^{-1},10]$.

Finally, we report in table \ref{tab:surveys} the details of the survey we analyse in section \ref{subsec:greffects} and~\ref{subsec:forecasts}, in particular the combinations of GWs observatories and large scale structure surveys, the covered fraction of the sky, the maximum multipole, connected to the maximum angular resolution achievable (see section \ref{sec:ggwcorrelation}) and the redshift binning we choose.

\begin{table}[h!]	
\centering
\begin{tabular}{|c|c|c|c|c|}
\hline
\makecell{SURVEYS  COMBINATION} & $f_{\mathrm{sky}}$ &  $\ell_{\mathrm{max}}$ & \makecell{REDSHIFT  BINNING} \\ \hline
GENERIC CASE  & 0.75 & 100 & \makecell{$1.0 \leq z_1 \leq 2.0$\\$2.0 \leq z_2 \leq 3.0$\\$ 3.0 \leq z_3 \leq 4.0$} \\ \hline
aLIGO $\times$ EMU  & 0.75 & 50 & \makecell{$0.0 \leq z_1 \leq 0.4$} \\ \hline
ET $\times$ DESI  & 0.34 & 100 & \makecell{$0.60 \leq z_1 \leq 1.15$\\$1.15 \leq z_2 \leq 1.70$} \\ \hline
ET $\times$ SKA  & 0.73 & 100 & \makecell{$1.0 \leq z_1 \leq 2.0$\\$2.0 \leq z_2 \leq 3.0$\\$ 3.0 \leq z_3 \leq 4.0$\\$4.0 \leq z_4 \leq 5.0$} \\ \hline
\end{tabular}
\caption{Prescription used for the forecast. We report the GWs observatory and the galaxy surveys, the covered fraction of the sky $f_{\mathrm{sky}}$, the maximum achievable multipole $\ell_{\mathrm{max}}$ of the GWs observatory and redshift binning. Note that we take the value $\ell_{\mathrm{max}}=50$ for aLIGO, because we assume that KAGRA and LIGO India will also be running, improving the source localisation and resolution of the resulting GWs events map.}
\label{tab:surveys}
\end{table}


\subsection{Generic Case and Importance of Projection Effects}
\label{subsec:greffects}
In this section we study a generic case to highlight the importance of projection effects. In terms of angular resolution and covered fraction of the sky, this generic case can be thought as a ET$\times$SKA in the redshift range $[1.0,4.0]$ (see table \ref{tab:surveys}).

\begin{figure}[h!]
\centerline{
\includegraphics[width=1.0\linewidth]{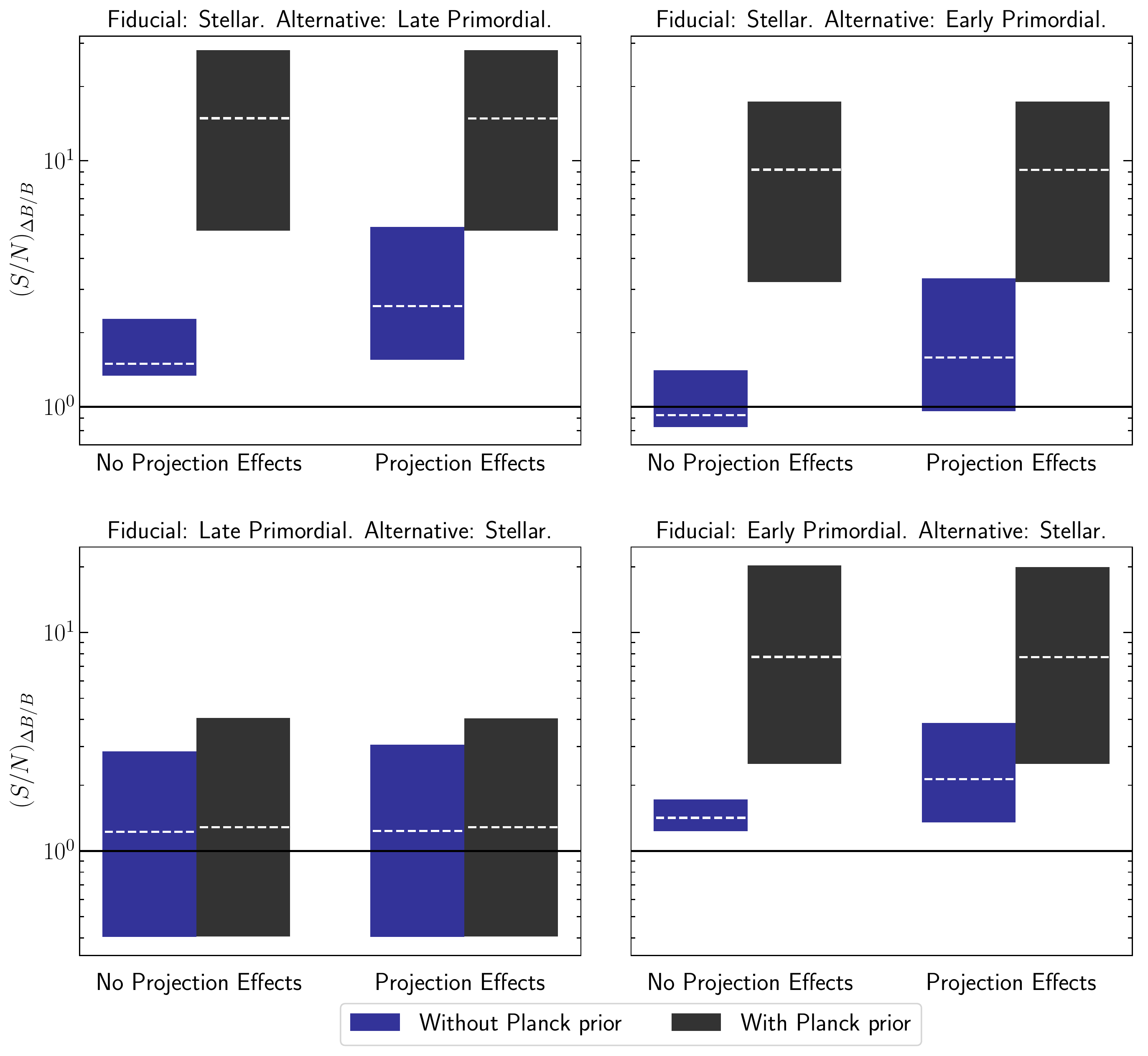}}
\caption{Generic case. Signal-to-Noise $(S/N)_{\Delta B/B}$ estimates from the Fisher analysis for different fiducial and alternative models, including and neglecting projection effects and the Planck prior on cosmological parameters. We assume $T_\mathrm{obs}=10$ years. The lower (upper) edge of the coloured bars corresponds to $r=0.1$ ($r=10$), while the white dashed line corresponds to the fiducial value $r=1$. The fiducial and the alternative models are indicated on top of each panel.}
\label{fig:SN_general}
\end{figure}

We perform the Fisher and $\Delta\chi^2$ analyses adding one by one the effects listed in equation \eqref{eq:numbercount_fluctuation} to estimate their importance, in particular we consider only density and velocity contributions ($\mathrm{den+vel}$ case), then we add lensing ($\mathrm{den+vel+len}$ case) and finally gravity effects ($\mathrm{den+vel+len+gr}$ case). In particular we call \textit{projection effects} the combination of the latter two, namely lensing and gravity contributions ($\mathrm{len+gr}$). We refer the interested reader to Refs. \cite{bellomo:multiclassI, bernal:multiclassII}, where a broader discussion on the importance of the full modelling can be found. Results from the Fisher analysis, with and without a Planck prior, are reported in figure \ref{fig:SN_general}, both for the stellar and primordial black holes scenarios. As expected, adding the CMB prior improves the results, lowering the value of the error $\sigma_{B^\mathrm{Fiducial}_{\mathrm{GW}}}$ defined in equation~\eqref{eq:SN_Fisher}, due to the extra power in constraining the standard cosmological parameters. We also find lensing effects have a large impact on determining the final Signal-to-Noise and that gravity contributions barely affects the final results, since they are relevant mostly at horizon scales.

We argue here that even if projection effects (lensing, gravity) do not depend on the signal one is trying to measure -- the bias in this case -- their contributions cannot necessarily be ignored in a Fisher error forecast for two reasons: {\it i)} they must be included in the covariance matrix as they act as an effective source of ``noise'' --~think of the cosmic variance contribution~-- as such ignoring them would underestimate the resulting error, {\it ii)} they do depend on other ``extra'' parameters, i.e. the cosmological parameters. When marginalizing over these extra parameters, the presence of projection effects help constraining them and thus improve the overall error-bars. The interplay and balance between these two trends yields a combined effect on the resulting forecasts, which we investigate now in more details.

\begin{figure}[h!]
\centerline{
\includegraphics[width=1.0\linewidth]{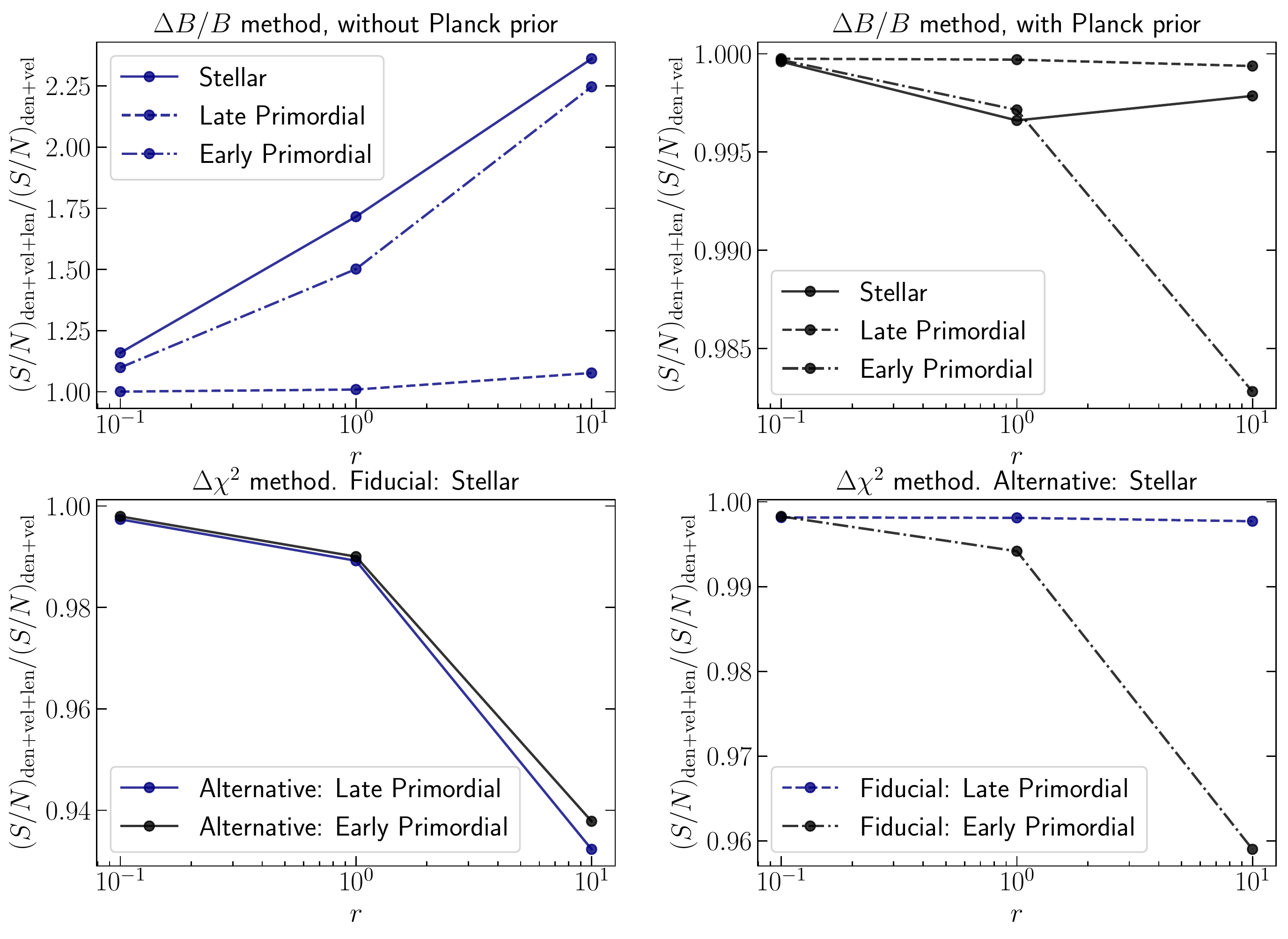}}
\caption{Generic case. \textit{Upper panels:} ratio of the Signal-to-Noise $(S/N)_{\Delta B/B}$ obtained with the Fisher analysis, including and neglecting projection effects, for different values of $r$, including (\textit{right panel}) or not (\textit{left panel}) the Planck prior. Lines indicates the assumed fiducial model. \textit{Lower panels:} ratio of the Signal-to-Noise $(S/N)_{\sqrt{\Delta\chi^2}}$ obtained with the $\Delta \chi^2$ analysis, including and neglecting projection effects, for different values of $r$, for the stellar as fiducial model (\textit{left panel}) or as alternative model (\textit{right panel}).}
\label{fig:ratio_lensing}
\end{figure}

We show on top panels of figure \ref{fig:ratio_lensing} how the inclusion of projection effects affects the results obtained through the Fisher analysis, depending on the value of the parameter $r$ and on the adoption of the Planck prior. In particular we appreciate that without adding Planck prior projection effects yield an improvement while adding it we observe a small degradation. This counter-intuitive result can be understood as follows. The lensing contribution does not directly depend on the bias parameter (see appendix \ref{app:relativistic_number_counts}) but it dominates the global $C_\ell$ signal, especially for the cross-bin angular power spectrum, acting as an effective source of ``noise'', since bigger $C_\ell$ yield smaller Fisher matrix elements (see equation \eqref{eq:Fisher}) and, consequently, a higher error $\sigma_{B^\mathrm{Fiducial}_\mathrm{GW}}$. On the other hand, lensing effects can improve forecasts on the other (cosmological) parameters considered in the Fisher analysis, increasing the corresponding Fisher elements. In the presence of degeneracies this can lead to an improvement on the bias parameter determination. This can be seen explicitly in the top left panel of figure \ref{fig:ratio_lensing}: in the three cases, for high enough values of $r$, the improvement of other cosmological parameters estimates breaks degeneracies, improving the forecasts on GWs bias $B_{\mathrm{GW}}$. This is not so evident for low values of the parameter $r$, as the higher shot noise works against the lensing-induced improvements on cosmological parameters. We have seen that in this method the inclusion of projection effects can change the forecast errors up to a factor of 2, therefore we argue that in general they cannot simply be neglected, even if there could be situations where the change is not so significant. When using the strong Planck prior, whose Fisher matrix elements are orders of magnitude bigger than those of the clustering, the lensing improvement on cosmological parameter forecasts is not significant any more, and we observed only the increased ``noise'' effect on $\sigma_{B^\mathrm{fiducial}_{\mathrm{GW}}}$.

In the lower panels of figure \ref{fig:ratio_lensing} we observe the same effect, this time with the $\Delta \chi^2$ method. In this case cosmological parameters are assumed to be known, and thus the lensing effect does not add signal, it only increases the noise.


\subsection{Forecast for Future Large Scale Structure Surveys}
\label{subsec:forecasts}
In this section we provide forecasts for those specific combinations of GWs observatories and large scale structure surveys given in table \ref{tab:surveys}.

\begin{figure}[h!]
\centerline{
\includegraphics[width=1.0\linewidth]{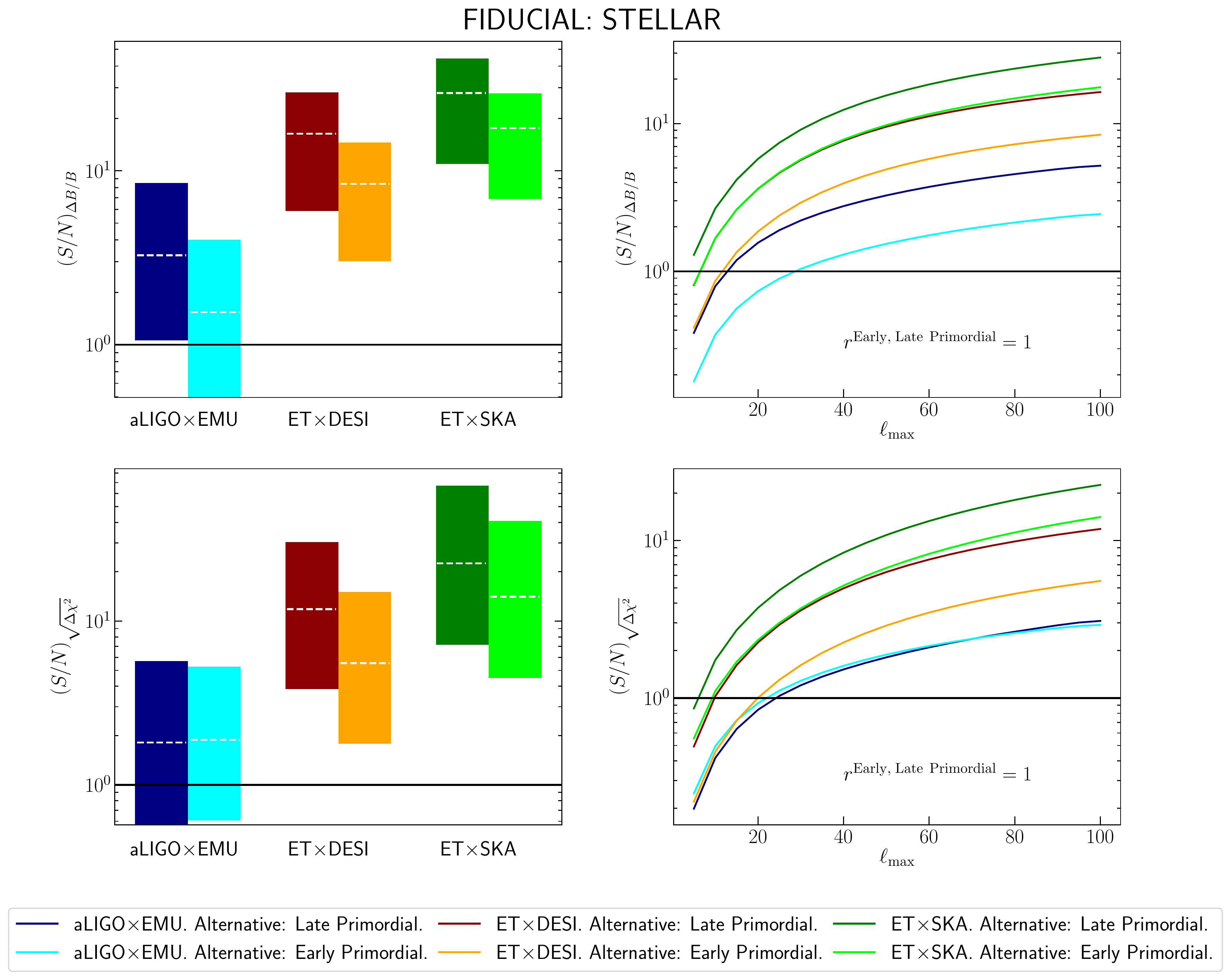}}
\caption{Specific surveys. Signal-to-Noise $S/N$ estimates coming from Fisher analyses, along with the Planck prior, (\textit{upper panels}) and $\Delta \chi^2$ formalism (\textit{lower panels}) for specific surveys combinations. \textit{Left panels:} Signal-to-Noise $S/N$ estimates as a function of $r$, assuming a fixed $\ell_{\mathrm{max}}$ ($50$ for aLIGO and $100$ for ET). The horizontal dashed white lines refer to the $r=1$ case. \textit{Right panels:} Signal-to-Noise $S/N$ estimates as a function of $\ell_{\mathrm{max}}$ for the fiducial merger rate case $r=1$. The fiducial scenario assumed is the stellar, to be distinguished by the early and late primordial alternative models. We choose as observation time $T_{\mathrm{obs}}=10$ years.}
\label{fig:SN_surveys_stellar}
\end{figure}

\begin{figure}[h!]
\centerline{
\includegraphics[width=1.0\linewidth]{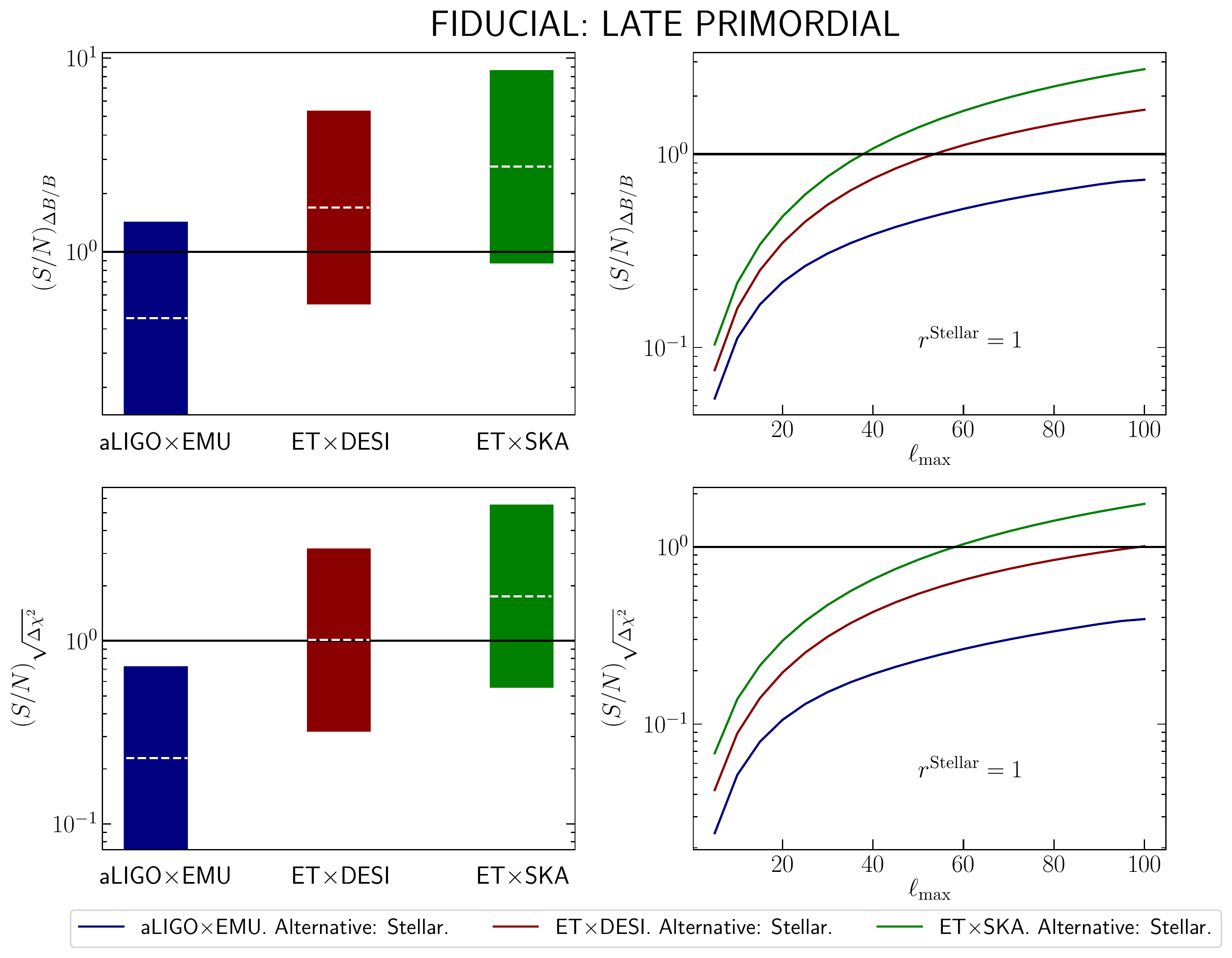}}
\caption{Specific surveys. Signal-to-Noise $S/N$ estimates coming from Fisher analyses, along with the Planck prior, (\textit{upper panels}) and $\Delta \chi^2$ formalism (\textit{lower panels}) for specific surveys combinations. \textit{Left panels:} Signal-to-Noise $S/N$ estimates as a function of $r$, assuming a fixed $\ell_{\mathrm{max}}$ ($50$ for aLIGO and $100$ for ET). The horizontal dashed white lines refer to the $r=1$ case. \textit{Right panels:} Signal-to-Noise $S/N$ estimates as a function of $\ell_{\mathrm{max}}$ for the fiducial merger rate case $r=1$. The horizontal dashed white lines refer to $r=1$. The fiducial scenario assumed is the late primordial, to be distinguished by the stellar model. We choose as observation time $T_{\mathrm{obs}}=10$ years.}
\label{fig:SN_surveys_late_primordial}
\end{figure}

\begin{figure}[h!]
\centerline{
\includegraphics[width=1.0\linewidth]{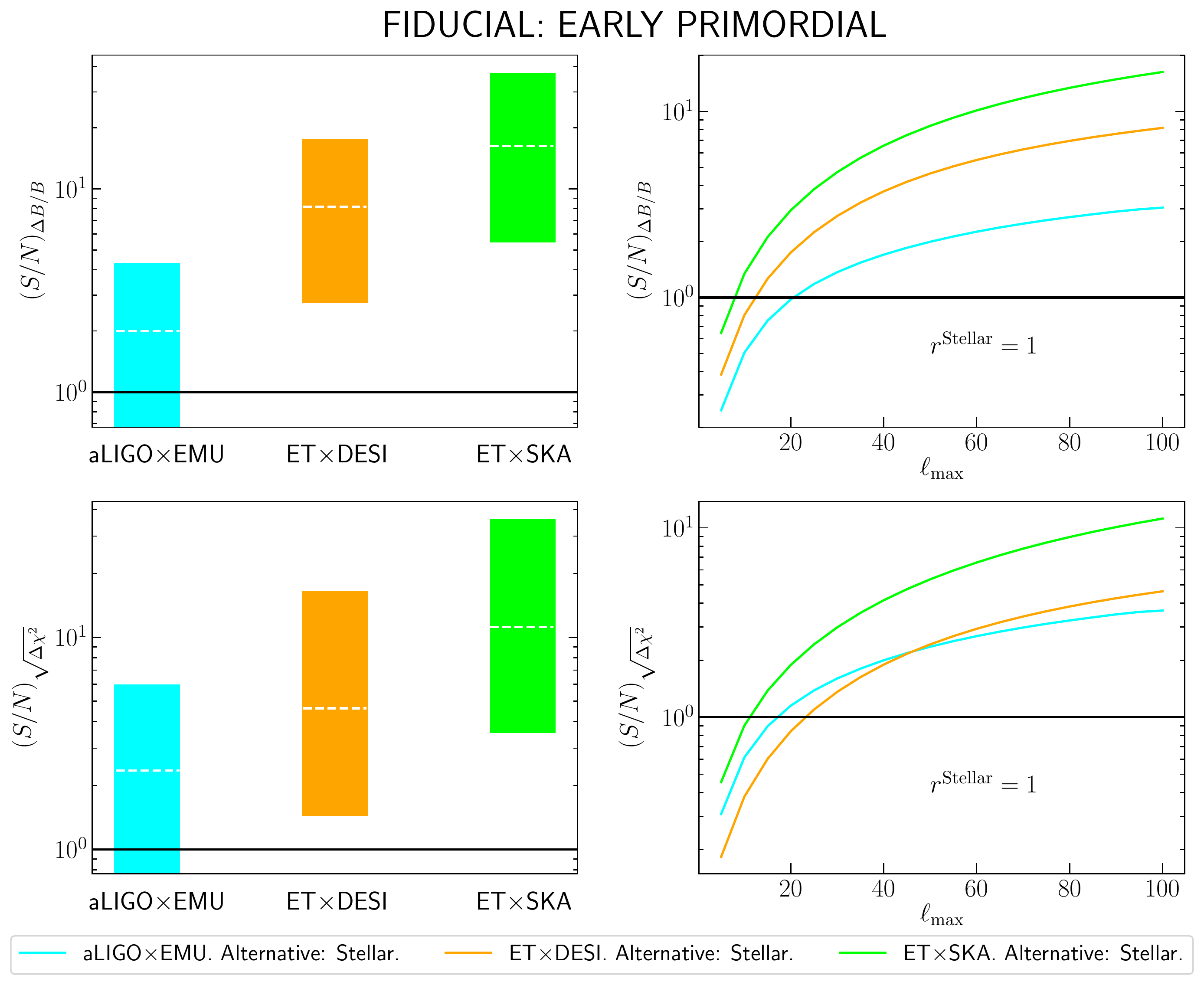}}
\caption{Specific surveys. Signal-to-Noise $S/N$ estimates coming from Fisher analyses, along with the Planck prior, (\textit{upper panels}) and $\Delta \chi^2$ formalism (\textit{lower panels}) for specific surveys combinations. \textit{Left panels:} Signal-to-Noise $S/N$ estimates as a function of $r$, assuming a fixed $\ell_{\mathrm{max}}$ ($50$ for aLIGO and $100$ for ET). The horizontal dashed white lines refer to the $r=1$ case. \textit{Right panels:} Signal-to-Noise $S/N$ estimates as a function of $\ell_{\mathrm{max}}$ for the fiducial merger rate case $r=1$. The fiducial scenario assumed is the early primordial, to be distinguished by the stellar model. We choose as observation time $T_{\mathrm{obs}}=10$ years.}
\label{fig:SN_surveys_early_primordial}
\end{figure}

We report the Signal-to-Noise forecasts, obtained with both methods described in section~\ref{sec:ggwcorrelation}, in figure~\ref{fig:SN_surveys_stellar} (stellar as fiducial model), figure~\ref{fig:SN_surveys_late_primordial} (late primordial as fiducial model) and figure~\ref{fig:SN_surveys_early_primordial} (early primordial as fiducial model). In each of these figures we show four panels: the upper ones show results coming from the Fisher analysis, while the lower ones come from the $\Delta \chi^2$ formalism. In the left panels we show bar charts obtained for different values of the parameter $r$ at fixed maximum multipole $\ell_{\mathrm{max}}$, while in the right panels we report the scaling of the Signal-to-Noise for different values of the maximum angular resolution when $r^\mathrm{Stellar, \: Late, \: Early\ Primordial}=1$, corresponding to the merger rates reported in equations~\eqref{eq:r_parameter} and~\eqref{eq:r_parameter_pbh}.

We show that surveys covering a bigger volume (or redshift range) have can discriminate better between different models, i.e. have higher Signal-to-Noise ratios, as expected from surveys with smaller shot noise. In the case where the stellar model is assumed as fiducial, it is generally more difficult to distinguish the early primordial scenario than the late primordial, since the former has a bias (or an effective bias) closer the stellar model one. Notice also that in the cases of stellar and early primordial as fiducial, we have better Signal-to-Noise ratio than in the late primordial scenario, due to higher merger rates, thus higher number of detected sources and lower shot noise.

In general we can conclude that future surveys will enable us to address questions about binary BHs mergers given enough observation time and resolution. One caveat is that this does not always happen for the aLIGO$\times$EMU combination, which will have a Signal-to-Noise lower or very close to unity in some cases (especially if mergers come from the late primordial formation mechanism). This is due to the fact that this combination of GWs observatory and large scale structure survey can only cover a low redshift range, where the biases (or the effective biases) are very similar (see e.g., the bottom left panel of Figure \ref{fig:dNdz_bias}) and we have an higher shot noise due to the scarce number of detected objects.


\section{Conclusions}
\label{sec:conclusions}
The renewed interest in primordial black holes has highlighted their importance not only as a possible constituent of the dark matter but also because their existence (if confirmed) would have profound implications about the physics of the early Universe. It is therefore essential to explore new ways to discriminate between primordial or stellar origin of the black holes which mergers have been observed with laser interferometers. Beyond the standard ways to constrain the existence of stellar mass primordial black holes through lensing or the effect on cosmic backgrounds, a complementary approach is to assess whether the GWs signal from merging binary BHs we detect are produced by objects of primordial origin or not.
 
Here we build on the idea that the cross-correlation of galaxy catalogues with GWs (from the merger of binary BHs) maps is a powerful tool to statistically study the origin of the progenitors of BHs mergers~\cite{raccanelli:pbhprogenitors}. This will be possible once the next generation of GWs detectors will provide localization of enough events to make low resolution maps. Galaxy catalogues covering a significant fraction of the sky and an overlapping redshift range are also under construction or at an advanced planning stage. Then, by measuring the bias of the halos hosting the binary BHs mergers, as well as the variation in their number counts due to lensing magnification and projection effects, we can infer the clustering properties of the progenitors of the binary BHs. Clustering properties matching those of luminous, high velocity-dispersion, high stellar-mass galaxies, would indicate a stellar origin, while clustering properties more similar to those of low-mass galaxies preferentially populating the filamentary structure of large-scale structures indicate a primordial origin. Moreover, this  approach could also set constraints on the abundance of PBHs, and hence on the fraction of dark matter that can be comprised of them.

We consider different models for the binary BHs formation, accretion mechanism, merger rate and clustering properties, both for the stellar and primordial nature of the BHs. We generalize similar studies on the cross-correlation between galaxy and gravitational wave maps by performing a full multi-tracer analysis that accounts for different redshift distributions, galaxy bias evolution, magnification bias of luminous sources as well as GWs, and relativistic projection effects. To perform such analyses one need to include a variety of different quantities and physical effects that are still poorly understood. For this reason, we investigated several possible scenarios and reported our results for a wide range of values of uncertain parameters, so that our results are quite general and can be still used once some of such quantities will be better understood.

Before studying specific forthcoming experiments, we highlighted the importance of projections effects on the estimate capability of a given experiment to differentiate between different origins of BHs mergers. Then, we find that the near future combination of aLIGO$\times$EMU would be able to address the nature of observed mergers in the case where their origin is mostly stellar or, if primordial, dominated by early binaries, due to a higher merger rate. If instead BHs have a late primordial origin, the corresponding merger rate would be too low to make it possible to distinguish this scenario from the stellar one with current observations. On the other hand, more futuristic survey combinations, such as ET$\times$DESI or ET$\times$SKA, would allow a real discrimination between all possible model combinations. Our results show that forthcoming experiments could allow us to test most of the parameter space of the still viable models investigated, and shed more light on the issue of binary black hole origin and evolution.

However, it is important to keep in mind some of the assumptions made in this work, that we will list below; it will therefore be important to keep refining these studies in order to have a robust model, formalism and pipeline once laser interferometers will deliver large GWs maps ready to be correlated with other datasets. Most of all, it is very difficult to understand, in the absence of full simulations, the process of binary formation and evolution of early binaries for PBHs scenarios. Related to this, another major source of uncertainty is the BHs accretion mechanism and efficiency across different cosmic epochs; estimates available in literature vary by several orders of magnitude.

In this work we considered PBHs with a monochromatic mass distribution peaked around $30 M_{\odot}$; while we verified that our results hold when considering instead some commonly used extended mass distributions, these types of analyses will differ when considering very different mass ranges. Finally, for the stellar model, uncertainties in the star formation rate, existence and distribution of sub-halos, massive star ejections and the epoch of first star formation can as well influence conclusions drawn from the GW$\times$LSS cross-correlation, and therefore need to be further studied. Nonetheless, we believe that the present work can contribute to further develop the new avenue of GW-LSS synergies, and that the vast range of parameters and models explored here make our results general enough to provide a realistic forecast of what this can teach us on the nature of binary BHs progenitors in the next decade.


\section*{Acknowledgments}
We thank Sathyaprakash Bangalore, Enrico Barausse, Jos\'e Luis Bernal, Anna Bonaldi, Yacine Ali-Ha\"imoud, Ely Kovetz, Julien Lesgourgues, Antonio Riotto and Matteo Viel for comments on the draft. We also thank Stefano Camera, Neal Dalal, Vincent Desjacques, Raul Jimenez and Sergey Sibiryakov for discussion. Funding for this work was partially provided by the Spanish MINECO under projects AYA2014-58747-P AEI/FEDER, UE, and MDM-2014-0369 of ICCUB (Unidad de Excelencia Mar\'ia de Maeztu). GS was supported by the Erasmus+ for Trainership grant during the early stages of this work, subsequently by grant from the ``Maria de Maeztu de Ci\`encies del Cosmos'' project mentioned above. NB is supported by the Spanish MINECO under grant BES-2015-073372. AR has received funding from the People Programme (Marie Curie Actions) of the European Union H2020 Programme under REA grant agreement number 706896 (COSMOFLAGS). SM acknowledges partial financial support by ASI Grant No. 2016-24-H.0. LV acknowledges support by European Union's Horizon 2020 research and innovation programme ERC (BePreSySe, grant agreement 725327). LV acknowledges the Radcliffe Institute for Advanced Study of Harvard University for hospitality during the latest stages of this work.


\bibliography{biblio}

\providecommand{\href}[2]{#2}\begingroup\raggedright\begin{thebibliography}{100}

\bibitem{abbott:firstligodetection}
The {\bfseries LIGO Scientific Collaboration and Virgo Collaboration}, B.~P.
  Abbott {\em et~al.}, ``Observation of Gravitational Waves from a Binary Black
  Hole Merger'', \href{http://dx.doi.org/10.1103/PhysRevLett.116.061102}{{\em
  Phys. Rev. Lett.} {\bfseries 116} (Feb, 2016) 061102},
  \href{http://arxiv.org/abs/1602.03837}{{\ttfamily arXiv:1602.03837}}.

\bibitem{abbott:firstligodetectionproperties}
The {\bfseries LIGO Scientific Collaboration and Virgo Collaboration}, B.~P.
  Abbott {\em et~al.}, ``Properties of the Binary Black Hole Merger GW150914'',
  \href{http://dx.doi.org/10.1103/PhysRevLett.116.241102}{{\em Phys. Rev.
  Lett.} {\bfseries 116} (Jun, 2016) 241102},
  \href{http://arxiv.org/abs/1602.03840}{{\ttfamily arXiv:1602.03840}}.

\bibitem{belczynski:massivebhsmergers}
K.~Belczynski, M.~Dominik, T.~Bulik, R.~O'Shaughnessy, C.~Fryer, and D.~E.
  Holz, ``The Effect of Metallicity on the Detection Prospects for
  Gravitational Waves'',
  \href{http://dx.doi.org/10.1088/2041-8205/715/2/L138}{{\em The Astrophysical
  Journal Letters} {\bfseries 715} no.~2, (2010) L138},
  \href{http://arxiv.org/abs/1004.0386}{{\ttfamily arXiv:1004.0386}}.

\bibitem{dominik:massivebhsmergersI}
M.~Dominik, K.~Belczynski, C.~Fryer, D.~E. Holz, E.~Berti, T.~Bulik, I.~Mandel,
  and R.~O'Shaughnessy, ``Double Compact Objects. I. The Significance of the
  Common Envelope on Merger Rates'',
  \href{http://dx.doi.org/10.1088/0004-637X/759/1/52}{{\em The Astrophysical
  Journal} {\bfseries 759} no.~1, (2012) 52},
  \href{http://arxiv.org/abs/1202.4901}{{\ttfamily arXiv:1202.4901}}.

\bibitem{dominik:massivebhsmergersII}
M.~Dominik, K.~Belczynski, C.~Fryer, D.~E. Holz, E.~Berti, T.~Bulik, I.~Mandel,
  and R.~O'Shaughnessy, ``Double Compact Objects. II. Cosmological Merger
  Rates'', \href{http://dx.doi.org/10.1088/0004-637X/779/1/72}{{\em The
  Astrophysical Journal} {\bfseries 779} no.~1, (2013) 72},
  \href{http://arxiv.org/abs/1308.1546}{{\ttfamily arXiv:1308.1546}}.

\bibitem{dominik:massivebhsmergersIII}
M.~Dominik, E.~Berti, R.~O'Shaughnessy, I.~Mandel, K.~Belczynski, C.~Fryer,
  D.~E. Holz, T.~Bulik, and F.~Pannarale, ``Double Compact Objects III:
  Gravitational-wave Detection Rates'',
  \href{http://dx.doi.org/10.1088/0004-637X/806/2/263}{{\em The Astrophysical
  Journal} {\bfseries 806} no.~2, (2015) 263},
  \href{http://arxiv.org/abs/1405.7016}{{\ttfamily arXiv:1405.7016}}.

\bibitem{abbott:secondligodetection}
The {\bfseries LIGO Scientific Collaboration and Virgo Collaboration}, B.~P.
  Abbott {\em et~al.}, ``GW151226: Observation of Gravitational Waves from a
  22-Solar-Mass Binary Black Hole Coalescence'',
  \href{http://dx.doi.org/10.1103/PhysRevLett.116.241103}{{\em Phys. Rev.
  Lett.} {\bfseries 116} (Jun, 2016) 241103},
  \href{http://arxiv.org/abs/1606.04855}{{\ttfamily arXiv:1606.04855}}.

\bibitem{abbott:thirdligodetection}
The {\bfseries LIGO Scientific Collaboration and Virgo Collaboration}, B.~P.
  Abbott {\em et~al.}, ``GW170104: Observation of a 50-Solar-Mass Binary Black
  Hole Coalescence at Redshift 0.2'',
  \href{http://dx.doi.org/10.1103/PhysRevLett.118.221101}{{\em Phys. Rev.
  Lett.} {\bfseries 118} (Jun, 2017) 221101},
  \href{http://arxiv.org/abs/1706.01812}{{\ttfamily arXiv:1706.01812}}.

\bibitem{abbott:fourthligodetection}
The {\bfseries LIGO Scientific Collaboration and Virgo Collaboration}, B.~P.
  Abbott {\em et~al.}, ``GW170608: Observation of a 19 Solar-mass Binary Black
  Hole Coalescence'', \href{http://dx.doi.org/10.3847/2041-8213/aa9f0c}{{\em
  The Astrophysical Journal Letters} {\bfseries 851} no.~2, (2017) L35},
  \href{http://arxiv.org/abs/arXiv:1711.05578}{{\ttfamily
  arXiv:arXiv:1711.05578}}.

\bibitem{abbott:fifthligodetection}
The {\bfseries LIGO Scientific Collaboration and Virgo Collaboration}, B.~P.
  Abbott {\em et~al.}, ``GW170814: A Three-Detector Observation of
  Gravitational Waves from a Binary Black Hole Coalescence'',
  \href{http://dx.doi.org/10.1103/PhysRevLett.119.141101}{{\em Phys. Rev.
  Lett.} {\bfseries 119} (Oct, 2017) 141101},
  \href{http://arxiv.org/abs/1709.09660}{{\ttfamily arXiv:1709.09660}}.

\bibitem{belczynski:highmassbhI}
K.~Belczynski, D.~E. Holz, T.~Bulik, and R.~O'Shaughnessy, ``The first
  gravitational-wave source from the isolated evolution of two stars in the
  40-100 solar mass range'', \href{http://dx.doi.org/10.1038/nature18322}{{\em
  Nature} {\bfseries 534} (2016) 512},
  \href{http://arxiv.org/abs/1602.04531}{{\ttfamily arXiv:1602.04531}}.

\bibitem{belczynski:highmassbhII}
K.~Belczynski, T.~Bulik, C.~L. Fryer, A.~Ruiter, F.~Valsecchi, J.~S. Vink, and
  J.~R. Hurley, ``On the Maximum Mass of Stellar Black Holes'',
  \href{http://dx.doi.org/10.1088/0004-637X/714/2/1217}{{\em The Astrophysical
  Journal} {\bfseries 714} no.~2, (2010) 1217},
  \href{http://arxiv.org/abs/0904.2784}{{\ttfamily arXiv:0904.2784}}.

\bibitem{belczynski:highmassbhIII}
K.~Belczynski, A.~Heger, W.~Gladysz, A.~J. Ruiter, S.~Woosley, G.~Wiktorowicz,
  H.-Y. Chen, T.~Bulik, R.~O'Shaughnessy, D.~E. Holz, C.~L. Fryer, and
  E.~Berti, ``The effect of pair-instability mass loss on black-hole mergers'',
  \href{http://dx.doi.org/10.1051/0004-6361/201628980}{{\em A\&A} {\bfseries
  594} (2016) A97}, \href{http://arxiv.org/abs/1607.03116}{{\ttfamily
  arXiv:1607.03116}}.

\bibitem{bird:pbhasdarkmatter}
S.~Bird, I.~Cholis, J.~B. Mu\~noz, Y.~Ali-Ha\"{\i}moud, M.~Kamionkowski, E.~D.
  Kovetz, A.~Raccanelli, and A.~G. Riess, ``Did LIGO Detect Dark Matter?'',
  \href{http://dx.doi.org/10.1103/PhysRevLett.116.201301}{{\em Phys. Rev.
  Lett.} {\bfseries 116} (May, 2016) 201301},
  \href{http://arxiv.org/abs/1603.00464}{{\ttfamily arXiv:1603.00464}}.

\bibitem{clesse:pbhmerging}
S.~Clesse and J.~Garc\'ia-Bellido, ``The clustering of massive Primordial Black
  Holes as Dark Matter: Measuring their mass distribution with advanced LIGO'',
  \href{http://dx.doi.org/https://doi.org/10.1016/j.dark.2016.10.002}{{\em
  Physics of the Dark Universe} {\bfseries 15} no.~Supplement C, (2017) 142 --
  147}, \href{http://arxiv.org/abs/1603.05234}{{\ttfamily arXiv:1603.05234}}.

\bibitem{Zeldovich:pbh}
Y.~B. Zel'dovich and I.~Novikov, ``The hypothesis of cores retarded during
  expansion and the hot cosmological model'',
  \href{http://dx.doi.org/1966AZh....43..758Z}{{\em Soviet Astronomy}
  {\bfseries 10} (1967) 602}.

\bibitem{Hawking:pbh1971}
S.~Hawking, ``Gravitationally Collapsed Objects of Very Low Mass'',
  \href{http://dx.doi.org/10.1093/mnras/152.1.75}{{\em Monthly Notices of the
  Royal Astronomical Society} {\bfseries 152} no.~1, (1971) 75--78}.

\bibitem{CarrHawking:pbh}
B.~J. Carr and S.~W. Hawking, ``Black Holes in the Early Universe'',
  \href{http://dx.doi.org/10.1093/mnras/168.2.399}{{\em Monthly Notices of the
  Royal Astronomical Society} {\bfseries 168} no.~2, (1974) 399--415}.

\bibitem{musco:pbh}
I.~Musco, J.~C. Miller, and L.~Rezzolla, ``Computations of primordial
  black-hole formation'',
  \href{http://dx.doi.org/10.1088/0264-9381/22/7/013}{{\em Classical and
  Quantum Gravity} {\bfseries 22} no.~7, (2005) 1405},
  \href{http://arxiv.org/abs/gr-qc/0412063}{{\ttfamily arXiv:gr-qc/0412063}}.

\bibitem{Polnarev88}
A.~Polnarev and R.~Zembowicz, ``Formation of primordial black holes by cosmic
  strings'', \href{http://dx.doi.org/10.1103/PhysRevD.43.1106}{{\em Phys. Rev.
  D} {\bfseries 43} (Feb, 1991) 1106--1109}.

\bibitem{HAWKING1989237}
S.~Hawking, ``Black holes from cosmic strings'',
  \href{http://dx.doi.org/https://doi.org/10.1016/0370-2693(89)90206-2}{{\em
  Physics Letters B} {\bfseries 231} no.~3, (1989) 237 -- 239}.

\bibitem{WICHOSKI1998191}
U.~F. Wichoski, J.~H. MacGibbon, and R.~H. Brandenberger, ``Astrophysical
  constraints on primordial black hole formation from collapsing cosmic
  strings'', \href{http://dx.doi.org/10.1016/S0370-1573(98)00070-2}{{\em
  Physics Reports} {\bfseries 307} no.~1, (1998) 191 -- 196}.

\bibitem{BEREZIN198391}
V.~Berezin, V.~Kuzmin, and I.~Tkachev, ``Thin-wall vacuum domain evolution'',
  \href{http://dx.doi.org/https://doi.org/10.1016/0370-2693(83)90630-5}{{\em
  Physics Letters B} {\bfseries 120} no.~1, (1983) 91 -- 96}.

\bibitem{Ipser84}
J.~Ipser and P.~Sikivie, ``Gravitationally repulsive domain wall'',
  \href{http://dx.doi.org/10.1103/PhysRevD.30.712}{{\em Phys. Rev. D}
  {\bfseries 30} (Aug, 1984) 712--719}.

\bibitem{Crawford82}
M.~Crawford and D.~N. Schramm, ``Spontaneous generation of density
  perturbations in the early Universe'',
  \href{http://dx.doi.org/http://dx.doi.org/10.1038/298538a0}{{\em Nature}
  {\bfseries 298} (1982) 538--540}.

\bibitem{LA1989375}
D.~La and P.~J. Steinhardt, ``Bubble percolation in extended inflationary
  models'',
  \href{http://dx.doi.org/https://doi.org/10.1016/0370-2693(89)90890-3}{{\em
  Physics Letters B} {\bfseries 220} no.~3, (1989) 375 -- 378}.

\bibitem{ivanov:pbhfrominflationI}
P.~Ivanov, P.~Naselsky, and I.~Novikov, ``Inflation and primordial black holes
  as dark matter'', \href{http://dx.doi.org/10.1103/PhysRevD.50.7173}{{\em
  Phys. Rev. D} {\bfseries 50} (Dec, 1994) 7173--7178}.

\bibitem{ivanov:pbhfrominflationII}
P.~Ivanov, ``Nonlinear metric perturbations and production of primordial black
  holes'', \href{http://dx.doi.org/10.1103/PhysRevD.57.7145}{{\em Phys. Rev. D}
  {\bfseries 57} (Jun, 1998) 7145--7154},
  \href{http://arxiv.org/abs/astro-ph/9708224}{{\ttfamily
  arXiv:astro-ph/9708224}}.

\bibitem{Bellido:pbh}
J.~Garc\'{\i}a-Bellido, A.~Linde, and D.~Wands, ``Density perturbations and
  black hole formation in hybrid inflation'',
  \href{http://dx.doi.org/10.1103/PhysRevD.54.6040}{{\em Phys. Rev. D}
  {\bfseries 54} (Nov, 1996) 6040--6058},
  \href{http://arxiv.org/abs/astro-ph/9605094}{{\ttfamily
  arXiv:astro-ph/9605094}}.

\bibitem{shandera:pbhfromdarkmatter}
S.~Shandera, D.~Jeong, and H.~S.~G. Gebhardt, ``Gravitational Waves from Binary
  Mergers of Subsolar Mass Dark Black Holes'',
  \href{http://dx.doi.org/10.1103/PhysRevLett.120.241102}{{\em Phys. Rev.
  Lett.} {\bfseries 120} (Jun, 2018) 241102},
  \href{http://arxiv.org/abs/1802.08206}{{\ttfamily arXiv:1802.08206}}.

\bibitem{barnacka:femtolensingconstraint}
A.~Barnacka, J.-F. Glicenstein, and R.~Moderski, ``New constraints on
  primordial black holes abundance from femtolensing of gamma-ray bursts'',
  \href{http://dx.doi.org/10.1103/PhysRevD.86.043001}{{\em Phys. Rev. D}
  {\bfseries 86} (Aug, 2012) 043001},
  \href{http://arxiv.org/abs/1204.2056}{{\ttfamily arXiv:1204.2056}}.

\bibitem{katz:femtolensingconstraint}
A.~Katz, J.~Kopp, S.~Sibiryakov, and W.~Xue, ``Femtolensing by Dark Matter
  Revisited'', \href{http://arxiv.org/abs/1807.11495}{{\ttfamily
  arXiv:1807.11495}}.

\bibitem{griest:keplerconstraint}
K.~Griest, A.~M. Cieplak, and M.~J. Lehner, ``Experimental Limits on Primordial
  Black Hole Dark Matter from the First 2 yr of Kepler Data'',
  \href{http://dx.doi.org/10.1088/0004-637X/786/2/158}{{\em The Astrophysical
  Journal} {\bfseries 786} no.~2, (2014) 158},
  \href{http://arxiv.org/abs/1307.5798}{{\ttfamily arXiv:1307.5798}}.

\bibitem{niikura:microlensingconstraint}
H.~Niikura, M.~Takada, N.~Yasuda, R.~H. Lupton, T.~Sumi, S.~More, A.~More,
  M.~Oguri, and M.~Chiba, ``Microlensing constraints on $10^{-10}M_\odot$-scale
  primordial black holes from high-cadence observation of M31 with Hyper
  Suprime-Cam'', \href{http://arxiv.org/abs/1701.02151}{{\ttfamily
  arXiv:1701.02151}}.

\bibitem{tisserand:microlensingconstraint}
The {\bfseries EROS-2 Collaboration}, P.~Tisserand {\em et~al.}, ``Limits on
  the Macho content of the Galactic Halo from the EROS-2 Survey of the
  Magellanic Clouds'', \href{http://dx.doi.org/10.1051/0004-6361:20066017}{{\em
  A\&A} {\bfseries 469} no.~2, (2007) 387--404},
  \href{http://arxiv.org/abs/astro-ph/0607207}{{\ttfamily
  arXiv:astro-ph/0607207}}.

\bibitem{calchinovati:microlensingconstraint}
S.~Calchi~Novati, S.~Mirzoyan, P.~Jetzer, and G.~Scarpetta, ``Microlensing
  towards the SMC: a new analysis of OGLE and EROS results'',
  \href{http://dx.doi.org/10.1093/mnras/stt1402}{{\em Monthly Notices of the
  Royal Astronomical Society} {\bfseries 435} no.~2, (2013) 1582--1597},
  \href{http://arxiv.org/abs/1308.4281}{{\ttfamily arXiv:1308.4281}}.

\bibitem{alcock:microlensingconstraint}
The {\bfseries MACHO Collaboration}, C.~Alcock {\em et~al.}, ``MACHO Project
  Limits on Black Hole Dark Matter in the 1-30 M$_\odot$ Range'',
  \href{http://dx.doi.org/10.1086/319636}{{\em The Astrophysical Journal
  Letters} {\bfseries 550} no.~2, (2001) L169},
  \href{http://arxiv.org/abs/astro-ph/0011506}{{\ttfamily
  arXiv:astro-ph/0011506}}.

\bibitem{mediavilla:microlensingconstraint}
E.~Mediavilla, J.~A. Munoz, E.~Falco, V.~Motta, E.~Guerras, H.~Canovas,
  C.~Jean, A.~Oscoz, and A.~M. Mosquera, ``Microlensing-based Estimate of the
  Mass Fraction in Compact Objects in Lens Galaxies'',
  \href{http://dx.doi.org/10.1088/0004-637X/706/2/1451}{{\em The Astrophysical
  Journal} {\bfseries 706} no.~2, (2009) 1451},
  \href{http://arxiv.org/abs/0910.3645}{{\ttfamily arXiv:0910.3645}}.

\bibitem{wilkinson:millilensingconstraint}
P.~N. Wilkinson, D.~R. Henstock, I.~W.~A. Browne, A.~G. Polatidis, P.~Augusto,
  A.~C.~S. Readhead, T.~J. Pearson, W.~Xu, G.~B. Taylor, and R.~C. Vermeulen,
  ``Limits on the Cosmological Abundance of Supermassive Compact Objects from a
  Search for Multiple Imaging in Compact Radio Sources'',
  \href{http://dx.doi.org/10.1103/PhysRevLett.86.584}{{\em Phys. Rev. Lett.}
  {\bfseries 86} (Jan, 2001) 584--587},
  \href{http://arxiv.org/abs/astro-ph/0101328}{{\ttfamily
  arXiv:astro-ph/0101328}}.

\bibitem{zumalacarregui:supernovaconstraint}
M.~Zumalacarregui and U.~Seljak, ``Limits on stellar-mass compact objects as
  dark matter from gravitational lensing of type Ia supernovae'',
  \href{http://arxiv.org/abs/1712.02240}{{\ttfamily arXiv:1712.02240}}.

\bibitem{graham:whitedwarfconstraint}
P.~W. Graham, S.~Rajendran, and J.~Varela, ``Dark matter triggers of
  supernovae'', \href{http://dx.doi.org/10.1103/PhysRevD.92.063007}{{\em Phys.
  Rev. D} {\bfseries 92} (Sep, 2015) 063007},
  \href{http://arxiv.org/abs/1505.04444}{{\ttfamily arXiv:1505.04444}}.

\bibitem{capela:neutronstarconstaint}
F.~Capela, M.~Pshirkov, and P.~Tinyakov, ``Constraints on primordial black
  holes as dark matter candidates from capture by neutron stars'',
  \href{http://dx.doi.org/10.1103/PhysRevD.87.123524}{{\em Phys. Rev. D}
  {\bfseries 87} (Jun, 2013) 123524},
  \href{http://arxiv.org/abs/1301.4984}{{\ttfamily arXiv:1301.4984}}.

\bibitem{quinn:widebinaryconstraint}
D.~P. Quinn, M.~I. Wilkinson, M.~J. Irwin, J.~Marshall, A.~Koch, and
  V.~Belokurov, ``On the reported death of the MACHO era'',
  \href{http://dx.doi.org/10.1111/j.1745-3933.2009.00652.x}{{\em Monthly
  Notices of the Royal Astronomical Society: Letters} {\bfseries 396} no.~1,
  (2009) L11--L15}, \href{http://arxiv.org/abs/0903.1644}{{\ttfamily
  arXiv:0903.1644}}.

\bibitem{brandt:ufdgconstraint}
T.~D. Brandt, ``Constraints on MACHO Dark Matter from Compact Stellar Systems
  in Ultra-faint Dwarf Galaxies'',
  \href{http://dx.doi.org/10.3847/2041-8205/824/2/L31}{{\em The Astrophysical
  Journal Letters} {\bfseries 824} no.~2, (2016) L31},
  \href{http://arxiv.org/abs/1605.03665}{{\ttfamily arXiv:1605.03665}}.

\bibitem{alihaimoud:pbhmergerrate}
Y.~Ali-Ha\"{\i}moud, E.~D. Kovetz, and M.~Kamionkowski, ``Merger rate of
  primordial black-hole binaries'',
  \href{http://dx.doi.org/10.1103/PhysRevD.96.123523}{{\em Phys. Rev. D}
  {\bfseries 96} (Dec, 2017) 123523},
  \href{http://arxiv.org/abs/1709.06576}{{\ttfamily arXiv:1709.06576}}.

\bibitem{magee:mergerrate}
R.~Magee, A.-S. Deutsch, P.~McClincy, C.~Hanna, C.~Horst, D.~Meacher,
  C.~Messick, S.~Shandera, and M.~Wade, ``Methods for the detection of
  gravitational waves from sub-solar mass ultracompact binaries'',
  \href{http://arxiv.org/abs/1808.04772}{{\ttfamily arXiv:1808.04772}}.

\bibitem{gaggero:accretionconstraints}
D.~Gaggero, G.~Bertone, F.~Calore, R.~M.~T. Connors, M.~Lovell, S.~Markoff, and
  E.~Storm, ``Searching for Primordial Black Holes in the radio and X-ray
  sky'', \href{http://dx.doi.org/10.1103/PhysRevLett.118.241101}{{\em Phys.
  Rev. Lett.} {\bfseries 118} no.~24, (2017) 241101},
  \href{http://arxiv.org/abs/1612.00457}{{\ttfamily arXiv:1612.00457}}.

\bibitem{ricotti:cmbconstraint}
M.~Ricotti, J.~P. Ostriker, and K.~J. Mack, ``Effect of Primordial Black Holes
  on the Cosmic Microwave Background and Cosmological Parameter Estimates'',
  \href{http://dx.doi.org/10.1086/587831}{{\em The Astrophysical Journal}
  {\bfseries 680} no.~2, (2008) 829},
  \href{http://arxiv.org/abs/0709.0524}{{\ttfamily arXiv:0709.0524}}.

\bibitem{alihamoud:pbhaccretion}
Y.~Ali-Ha\"{\i}moud and M.~Kamionkowski, ``Cosmic microwave background limits
  on accreting primordial black holes'',
  \href{http://dx.doi.org/10.1103/PhysRevD.95.043534}{{\em Phys. Rev. D}
  {\bfseries 95} (Feb, 2017) 043534},
  \href{http://arxiv.org/abs/1612.05644}{{\ttfamily arXiv:1612.05644}}.

\bibitem{poulin:cmbconstraint}
V.~Poulin, P.~D. Serpico, F.~Calore, S.~Clesse, and K.~Kohri, ``CMB bounds on
  disk-accreting massive primordial black holes'',
  \href{http://dx.doi.org/10.1103/PhysRevD.96.083524}{{\em Phys. Rev. D}
  {\bfseries 96} (Oct, 2017) 083524},
  \href{http://arxiv.org/abs/1707.04206}{{\ttfamily arXiv:1707.04206}}.

\bibitem{bernal:cmbconstraint}
J.~L. Bernal, N.~Bellomo, A.~Raccanelli, and L.~Verde, ``Cosmological
  Implications of Primordial Black Holes'',
  \href{http://dx.doi.org/10.1088/1475-7516/2017/10/052}{{\em JCAP} {\bfseries
  2017} no.~10, (2017) 052}, \href{http://arxiv.org/abs/1709.07465}{{\ttfamily
  arXiv:1709.07465}}.

\bibitem{KfirBlum}
D.~Aloni, K.~Blum, and R.~Flauger, ``Cosmic microwave background constraints on
  primordial black hole dark matter'',
  \href{http://dx.doi.org/10.1088/1475-7516/2017/05/017}{{\em Journal of
  Cosmology and Astroparticle Physics} {\bfseries 2017} no.~05, (2017) 017},
  \href{http://arxiv.org/abs/1612.06811}{{\ttfamily arXiv:1612.06811}}.

\bibitem{bellomo:emdconstraints}
N.~Bellomo, J.~L. Bernal, A.~Raccanelli, and L.~Verde, ``Primordial black holes
  as dark matter: converting constraints from monochromatic to extended mass
  distributions'', \href{http://dx.doi.org/10.1088/1475-7516/2018/01/004}{{\em
  Journal of Cosmology and Astroparticle Physics} {\bfseries 2018} no.~01,
  (2018) 004}, \href{http://arxiv.org/abs/1709.07467}{{\ttfamily
  arXiv:1709.07467}}.

\bibitem{ALIGO:aligo}
The {\bfseries LIGO Scientific Collaboration}, ``Advanced LIGO'',
  \href{http://dx.doi.org/10.1088/0264-9381/32/7/074001}{{\em Classical and
  Quantum Gravity} {\bfseries 32} no.~7, (2015) 074001},
  \href{http://arxiv.org/abs/1411.4547}{{\ttfamily arXiv:1411.4547}}.

\bibitem{Sathyaprakash:ET}
B.~Sathyaprakash {\em et~al.}, ``Scientific objectives of Einstein Telescope'',
  \href{http://dx.doi.org/10.1088/0264-9381/29/12/124013}{{\em Classical and
  Quantum Gravity} {\bfseries 29} no.~12, (2012) 124013},
  \href{http://arxiv.org/abs/1206.0331}{{\ttfamily arXiv:1206.0331}}.

\bibitem{raccanelli:pbhprogenitors}
A.~Raccanelli, E.~D. Kovetz, S.~Bird, I.~Cholis, and J.~B. Mu\~noz,
  ``Determining the progenitors of merging black-hole binaries'',
  \href{http://dx.doi.org/10.1103/PhysRevD.94.023516}{{\em Phys. Rev. D}
  {\bfseries 94} (Jul, 2016) 023516},
  \href{http://arxiv.org/abs/1605.01405}{{\ttfamily arXiv:1605.01405}}.

\bibitem{raccanelli:gwastronomy}
A.~Raccanelli, ``Gravitational wave astronomy with radio galaxy surveys'',
  \href{http://dx.doi.org/10.1093/mnras/stx835}{{\em Monthly Notices of the
  Royal Astronomical Society} {\bfseries 469} no.~1, (2017) 656--670},
  \href{http://arxiv.org/abs/1609.09377}{{\ttfamily arXiv:1609.09377}}.

\bibitem{nishikawa:pbhmergers}
H.~Nishikawa, E.~D. Kovetz, M.~Kamionkowski, and J.~Silk,
  ``Primordial-black-hole mergers in dark-matter spikes'',
  \href{http://arxiv.org/abs/1708.08449}{{\ttfamily arXiv:1708.08449}}.

\bibitem{cholis:orbitaleccentricities}
I.~Cholis, E.~D. Kovetz, Y.~Ali-Ha{\"\i}moud, S.~Bird, M.~Kamionkowski, J.~B.
  Mu{\~n}oz, and A.~Raccanelli, ``Orbital eccentricities in primordial black
  hole binaries'', \href{http://dx.doi.org/10.1103/PhysRevD.94.084013}{{\em
  Physical Review D} {\bfseries 94} no.~8, (2016) 084013},
  \href{http://arxiv.org/abs/1606.07437}{{\ttfamily arXiv:1606.07437}}.

\bibitem{munoz:fastradioburst}
J.~B. Mu{\~n}oz, E.~D. Kovetz, L.~Dai, and M.~Kamionkowski, ``Lensing of Fast
  Radio Bursts as a Probe of Compact Dark Matter'',
  \href{http://dx.doi.org/10.1103/PhysRevLett.117.091301}{{\em Phys. Rev.
  Lett.} {\bfseries 117} no.~9, (2016) 091301},
  \href{http://arxiv.org/abs/1605.00008}{{\ttfamily arXiv:1605.00008}}.

\bibitem{kovetz:pbhmassfunction}
E.~D. Kovetz, I.~Cholis, P.~C. Breysse, and M.~Kamionkowski, ``Black hole mass
  function from gravitational wave measurements'',
  \href{http://dx.doi.org/10.1103/PhysRevD.95.103010}{{\em Phys. Rev. D}
  {\bfseries 95} (May, 2017) 103010},
  \href{http://arxiv.org/abs/1611.01157}{{\ttfamily arXiv:1611.01157}}.

\bibitem{kovetz:pbhandgw}
E.~D. Kovetz, ``Probing Primordial Black Hole Dark Matter with Gravitational
  Waves'', \href{http://dx.doi.org/10.1103/PhysRevLett.119.131301}{{\em Phys.
  Rev. Lett.} {\bfseries 119} (Sep, 2017) 131301},
  \href{http://arxiv.org/abs/1705.09182}{{\ttfamily arXiv:1705.09182}}.

\bibitem{Munari:velocity_mass_halos}
E.~Munari, A.~Biviano, S.~Borgani, G.~Murante, and D.~Fabjan, ``The relation
  between velocity dispersion and mass in simulated clusters of galaxies:
  dependence on the tracer and the baryonic physics'',
  \href{http://dx.doi.org/10.1093/mnras/stt049}{{\em Monthly Notices of the
  Royal Astronomical Society} {\bfseries 430} no.~4, (2013) 2638--2649},
  \href{http://arxiv.org/abs/1301.1682}{{\ttfamily arXiv:1301.1682}}.

\bibitem{Vale:luminosity_mass_halos}
A.~Vale and J.~P. Ostriker, ``Linking halo mass to galaxy luminosity'',
  \href{http://dx.doi.org/10.1111/j.1365-2966.2004.08059.x}{{\em Monthly
  Notices of the Royal Astronomical Society} {\bfseries 353} no.~1, (2004)
  189--200}, \href{http://arxiv.org/abs/0402500}{{\ttfamily arXiv:0402500}}.

\bibitem{erb:starformationrate}
D.~K. Erb, ``Feedback in low-mass galaxies in the early Universe'',
  \href{http://dx.doi.org/10.1038/nature14454}{{\em Nature} {\bfseries 523}
  (Jul, 2015) 169--176}, \href{http://arxiv.org/abs/1507.02374}{{\ttfamily
  arXiv:1507.02374}}.

\bibitem{Norris:EMU}
R.~P. Norris {\em et~al.}, ``EMU: Evolutionary Map of the Universe'',
  \href{http://dx.doi.org/10.1071/AS11021}{{\em Publications of the
  Astronomical Society of Australia} {\bfseries 28} no.~3, (2011) 215--248},
  \href{http://arxiv.org/abs/1106.3219}{{\ttfamily arXiv:1106.3219}}.

\bibitem{Aghamousa:desi}
The {\bfseries DESI Collaboration}, A.~Aghamousa {\em et~al.}, ``The DESI
  Experiment Part I: Science, Targeting, and Survey Design'',
  \href{http://arxiv.org/abs/1611.00036}{{\ttfamily arXiv:1611.00036}}.

\bibitem{maartens:SKA}
R.~Maartens {\em et~al.}, ``Cosmology with the SKA - overview'',
  \href{http://arxiv.org/abs/1501.04076}{{\ttfamily arXiv:1501.04076}}.

\bibitem{dai:gravitationallensing}
L.~Dai, T.~Venumadhav, and K.~Sigurdson, ``Effect of lensing magnification on
  the apparent distribution of black hole mergers'',
  \href{http://dx.doi.org/10.1103/PhysRevD.95.044011}{{\em Phys. Rev. D}
  {\bfseries 95} (Feb, 2017) 044011},
  \href{http://arxiv.org/abs/1605.09398}{{\ttfamily arXiv:1605.09398}}.

\bibitem{oguri:gravitationallensing}
M.~Oguri, ``Effect of gravitational lensing on the distribution of
  gravitational waves from distant binary black hole mergers'',
  \href{http://dx.doi.org/10.1093/mnras/sty2145}{{\em Monthly Notices of the
  Royal Astronomical Society} {\bfseries 480} no.~3, (2018) 3842--3855},
  \href{http://arxiv.org/abs/1807.02584}{{\ttfamily arXiv:1807.02584}}.

\bibitem{wang:lensing}
Y.~Wang, A.~Stebbins, and E.~L. Turner, ``Gravitational lensing of
  gravitational waves from merging neutron star binaries'',
  \href{http://dx.doi.org/10.1103/PhysRevLett.77.2875}{{\em Physical review
  letters} {\bfseries 77} no.~14, (1996) 2875},
  \href{http://arxiv.org/abs/astro-ph/9605140}{{\ttfamily
  arXiv:astro-ph/9605140}}.

\bibitem{bertacca:luminositydistance}
D.~Bertacca, A.~Raccanelli, N.~Bartolo, and S.~Matarrese, ``Cosmological
  perturbation effects on gravitational-wave luminosity distance estimates'',
  \href{http://dx.doi.org/10.1016/j.dark.2018.03.001}{{\em Physics of the Dark
  Universe} {\bfseries 20} (2018) 32 -- 40},
  \href{http://arxiv.org/abs/1702.01750}{{\ttfamily arXiv:1702.01750}}.

\bibitem{camera:gwlensing}
S.~Camera and A.~Nishizawa, ``Beyond Concordance Cosmology with Magnification
  of Gravitational-Wave Standard Sirens'',
  \href{http://dx.doi.org/10.1103/PhysRevLett.110.151103}{{\em Phys. Rev.
  Lett.} {\bfseries 110} (Apr, 2013) 151103},
  \href{http://arxiv.org/abs/1303.5446}{{\ttfamily arXiv:1303.5446}}.

\bibitem{schutz:lmax}
B.~F. Schutz, ``Networks of gravitational wave detectors and three figures of
  merit'', \href{http://dx.doi.org/10.1088/0264-9381/28/12/125023}{{\em
  Classical and Quantum Gravity} {\bfseries 28} no.~12, (2011) 125023},
  \href{http://arxiv.org/abs/1102.5421}{{\ttfamily arXiv:1102.5421}}.

\bibitem{klimenko:lmax}
S.~Klimenko, G.~Vedovato, M.~Drago, G.~Mazzolo, G.~Mitselmakher, C.~Pankow,
  G.~Prodi, V.~Re, F.~Salemi, and I.~Yakushin, ``Localization of gravitational
  wave sources with networks of advanced detectors'',
  \href{http://dx.doi.org/10.1103/PhysRevD.83.102001}{{\em Physical Review D}
  {\bfseries 83} no.~10, (2011) 102001},
  \href{http://arxiv.org/abs/1101.5408}{{\ttfamily arXiv:1101.5408}}.

\bibitem{sidery:lmax}
T.~Sidery, B.~Aylott, N.~Christensen, B.~Farr, W.~Farr, F.~Feroz, J.~Gair,
  K.~Grover, P.~Graff, C.~Hanna, {\em et~al.}, ``Reconstructing the sky
  location of gravitational-wave detected compact binary systems: methodology
  for testing and comparison'',
  \href{http://dx.doi.org/10.1103/PhysRevD.89.084060}{{\em Physical Review D}
  {\bfseries 89} no.~8, (2014) 084060},
  \href{http://arxiv.org/abs/1312.6013}{{\ttfamily arXiv:1312.6013}}.

\bibitem{namikawa:lmax}
T.~Namikawa, A.~Nishizawa, and A.~Taruya, ``Anisotropies of gravitational-wave
  standard sirens as a new cosmological probe without redshift information'',
  \href{http://dx.doi.org/10.1103/PhysRevLett.116.121302}{{\em Physical review
  letters} {\bfseries 116} no.~12, (2016) 121302},
  \href{http://arxiv.org/abs/1511.04638}{{\ttfamily arXiv:1511.04638}}.

\bibitem{raccanelli:crosscorrelation}
A.~Raccanelli, A.~Bonaldi, M.~Negrello, S.~Matarrese, G.~Tormen, and
  G.~De~Zotti, ``A reassessment of the evidence of the Integrated Sachs-Wolfe
  effect through the WMAP-NVSS correlation'',
  \href{http://dx.doi.org/10.1111/j.1365-2966.2008.13189.x}{{\em Monthly
  Notices of the Royal Astronomical Society} {\bfseries 386} no.~4, (2008)
  2161--2166}, \href{http://arxiv.org/abs/0802.0084}{{\ttfamily
  arXiv:0802.0084}}.

\bibitem{pullen:crosscorrelation}
A.~R. Pullen, T.-C. Chang, O.~Dor\'{e}, and A.~Lidz, ``Cross-correlations as a
  Cosmological Carbon Monoxide Detector'',
  \href{http://dx.doi.org/10.1088/0004-637X/768/1/15}{{\em The Astrophysical
  Journal} {\bfseries 768} no.~1, (2013) 15},
  \href{http://arxiv.org/abs/1211.1397}{{\ttfamily arXiv:1211.1397}}.

\bibitem{bonvin:cl}
C.~Bonvin and R.~Durrer, ``What galaxy surveys really measure'',
  \href{http://dx.doi.org/10.1103/PhysRevD.84.063505}{{\em Phys. Rev. D}
  {\bfseries 84} (Sep, 2011) 063505},
  \href{http://arxiv.org/abs/1105.5280}{{\ttfamily arXiv:1105.5280}}.

\bibitem{blas:class}
D.~Blas, J.~Lesgourgues, and T.~Tram, ``The Cosmic Linear Anisotropy Solving
  System (CLASS). Part II: Approximation schemes'',
  \href{http://dx.doi.org/10.1088/1475-7516/2011/07/034}{{\em Journal of
  Cosmology and Astroparticle Physics} {\bfseries 2011} no.~07, (2011) 034},
  \href{http://arxiv.org/abs/1104.2933}{{\ttfamily arXiv:1104.2933}}.

\bibitem{didio:classgal}
E.~D. Dio, F.~Montanari, J.~Lesgourgues, and R.~Durrer, ``The CLASSgal code for
  relativistic cosmological large scale structure'',
  \href{http://dx.doi.org/10.1088/1475-7516/2013/11/044}{{\em Journal of
  Cosmology and Astroparticle Physics} {\bfseries 2013} no.~11, (2013) 044},
  \href{http://arxiv.org/abs/1307.1459}{{\ttfamily arXiv:1307.1459}}.

\bibitem{challinor:deltag}
A.~Challinor and A.~Lewis, ``Linear power spectrum of observed source number
  counts'', \href{http://dx.doi.org/10.1103/PhysRevD.84.043516}{{\em Phys. Rev.
  D} {\bfseries 84} (Aug, 2011) 043516},
  \href{http://arxiv.org/abs/1105.5292}{{\ttfamily arXiv:1105.5292}}.

\bibitem{bellomo:multiclassI}
N.~Bellomo, J.~L. Bernal, G.~Scelfo, A.~Raccanelli, and L.~Verde, ``Beware of
  commonly used approximations I: errors in forecasts'',
  \href{http://arxiv.org/abs/2005.10384}{{\ttfamily arXiv:2005.10384}}.

\bibitem{bernal:multiclassII}
J.~L. Bernal, N.~Bellomo, A.~Raccanelli, and L.~Verde, ``Beware of commonly
  used approximations II: estimating systematic biases in the best-fit
  parameters'', \href{http://arxiv.org/abs/2005.09666}{{\ttfamily
  arXiv:2005.09666}}.

\bibitem{unnikhrishnan:indigo}
C.~S. Unnikrishnan, ``IndIGO and LIGO-India: Scope and Plans for Gravitational
  Wave Research and Precision Metrology in India'',
  \href{http://dx.doi.org/10.1142/S0218271813410101}{{\em International Journal
  of Modern Physics D} {\bfseries 22} no.~01, (2013) 1341010},
  \href{http://arxiv.org/abs/1510.06059}{{\ttfamily arXiv:1510.06059}}.

\bibitem{somiya:kagra}
K.~Somiya, ``Detector configuration of KAGRA - the Japanese cryogenic
  gravitational-wave detector'',
  \href{http://dx.doi.org/10.1088/0264-9381/29/12/124007}{{\em Classical and
  Quantum Gravity} {\bfseries 29} no.~12, (2012) 124007},
  \href{http://arxiv.org/abs/1111.7185}{{\ttfamily arXiv:1111.7185}}.

\bibitem{Bunn:Fisher}
E.~F. Bunn, {\em Statistical analysis of cosmic microwave background
  anisotropy}.
\newblock PhD thesis, University of California, Berkeley, 1995.

\bibitem{Tegmark:Fisher}
M.~Tegmark, A.~N. Taylor, and A.~F. Heavens, ``Karhunen-Lo\`eve Eigenvalue
  Problems in Cosmology: How Should We Tackle Large Data Sets?'',
  \href{http://dx.doi.org/10.1086/303939}{{\em The Astrophysical Journal}
  {\bfseries 480} no.~1, (1997) 22},
  \href{http://arxiv.org/abs/astro-ph/9603021}{{\ttfamily
  arXiv:astro-ph/9603021}}.

\bibitem{Vogeley:FIsher}
M.~S. Vogeley and A.~S. Szalay, ``Eigenmode analysis of galaxy redshift surveys
  I. theory and methods'', \href{http://dx.doi.org/10.1086/177399}{{\em The
  Astrophysical Journal} {\bfseries 465} (1996) 34},
  \href{http://arxiv.org/abs/astro-ph/9601185}{{\ttfamily
  arXiv:astro-ph/9601185}}.

\bibitem{Planck:XIII}
The {\bfseries Planck Collaboration}, P.~A.~R. Ade {\em et~al.}, ``Planck 2015
  results - XIII. Cosmological parameters'',
  \href{http://dx.doi.org/10.1051/0004-6361/201525830}{{\em A\&A} {\bfseries
  594} (2016) A13}, \href{http://arxiv.org/abs/1502.01589}{{\ttfamily
  arXiv:1502.01589}}.

\bibitem{jarvis:sfgnumberdensity}
M.~J. Jarvis, D.~Bacon, C.~Blake, M.~L. Brown, S.~N. Lindsay, A.~Raccanelli,
  M.~Santos, and D.~Schwarz, ``Cosmology with SKA Radio Continuum Surveys'',
  \href{http://arxiv.org/abs/1501.03825}{{\ttfamily arXiv:1501.03825}}.

\bibitem{Bonaldi:TRECS}
A.~Bonaldi, M.~Bonato, V.~Galluzzi, I.~Harrison, M.~Massardi, S.~Kay,
  G.~De~Zotti, and M.~L. Brown, ``The Tiered Radio Extragalactic Continuum
  Simulation (T-RECS)'', \href{http://arxiv.org/abs/1805.05222}{{\ttfamily
  arXiv:1805.05222}}.

\bibitem{raccanelli:sfgbias}
A.~Raccanelli, G.-B. Zhao, D.~J. Bacon, M.~J. Jarvis, W.~J. Percival, R.~P.
  Norris, H.~R\"ottgering, F.~B. Abdalla, C.~M. Cress, J.-C. Kubwimana,
  S.~Lindsay, R.~C. Nichol, M.~G. Santos, and D.~J. Schwarz, ``Cosmological
  measurements with forthcoming radio continuum surveys'',
  \href{http://dx.doi.org/10.1111/j.1365-2966.2012.20634.x}{{\em Monthly
  Notices of the Royal Astronomical Society} {\bfseries 424} no.~2, (2012)
  801--819}, \href{http://arxiv.org/abs/1108.0930}{{\ttfamily
  arXiv:1108.0930}}.

\bibitem{wilman:radiosourcessimulation}
R.~J. Wilman, L.~Miller, M.~J. Jarvis, T.~Mauch, F.~Levrier, F.~B. Abdalla,
  S.~Rawlings, H.-R. Kl\"ockner, D.~Obreschkow, D.~Olteanu, and S.~Young, ``A
  semi-empirical simulation of the extragalactic radio continuum sky for next
  generation radio telescopes'',
  \href{http://dx.doi.org/10.1111/j.1365-2966.2008.13486.x}{{\em Monthly
  Notices of the Royal Astronomical Society} {\bfseries 388} no.~3, (2008)
  1335--1348}, \href{http://arxiv.org/abs/0805.3413}{{\ttfamily
  arXiv:0805.3413}}.

\bibitem{turner:magnificationbias}
E.~L. Turner, J.~P. Ostriker, and J.~R. Gott, III, ``The statistics of
  gravitational lenses - The distributions of image angular separations and
  lens redshifts'', \href{http://dx.doi.org/10.1086/162379}{{\em Astrophysical
  Journal} {\bfseries 284} (Sep, 1984) 1--22}.

\bibitem{bartelmann:convergence}
M.~Bartelmann and P.~Schneider, ``Weak gravitational lensing'',
  \href{http://dx.doi.org/https://doi.org/10.1016/S0370-1573(00)00082-X}{{\em
  Physics Reports} {\bfseries 340} no.~4, (2001) 291 -- 472},
  \href{http://arxiv.org/abs/astro-ph/9912508}{{\ttfamily
  arXiv:astro-ph/9912508}}.

\bibitem{hui:magnificationbias}
L.~Hui, E.~Gazta\~naga, and M.~LoVerde, ``Anisotropic magnification distortion
  of the 3D galaxy correlation. I. Real space'',
  \href{http://dx.doi.org/10.1103/PhysRevD.76.103502}{{\em Phys. Rev. D}
  {\bfseries 76} (Nov, 2007) 103502},
  \href{http://arxiv.org/abs/0706.1071}{{\ttfamily arXiv:0706.1071}}.

\bibitem{liu:magnificationbias}
J.~Liu, Z.~Haiman, L.~Hui, J.~M. Kratochvil, and M.~May, ``Impact of
  magnification and size bias on the weak lensing power spectrum and peak
  statistics'', \href{http://dx.doi.org/10.1103/PhysRevD.89.023515}{{\em Phys.
  Rev. D} {\bfseries 89} (Jan, 2014) 023515},
  \href{http://arxiv.org/abs/1310.7517}{{\ttfamily arXiv:1310.7517}}.

\bibitem{montanari:magnificationbias}
F.~Montanari and R.~Durrer, ``Measuring the lensing potential with tomographic
  galaxy number counts'',
  \href{http://dx.doi.org/10.1088/1475-7516/2015/10/070}{{\em Journal of
  Cosmology and Astroparticle Physics} {\bfseries 2015} no.~10, (2015) 070},
  \href{http://arxiv.org/abs/1506.01369}{{\ttfamily arXiv:1506.01369}}.

\bibitem{challinor:evolutionbias}
A.~Challinor and A.~Lewis, ``Linear power spectrum of observed source number
  counts'', \href{http://dx.doi.org/10.1103/PhysRevD.84.043516}{{\em Phys. Rev.
  D} {\bfseries 84} (Aug, 2011) 043516},
  \href{http://arxiv.org/abs/1105.5292}{{\ttfamily arXiv:1105.5292}}.

\bibitem{jeong:evolutionbias}
D.~Jeong, F.~Schmidt, and C.~M. Hirata, ``Large-scale clustering of galaxies in
  general relativity'',
  \href{http://dx.doi.org/10.1103/PhysRevD.85.023504}{{\em Phys. Rev. D}
  {\bfseries 85} (Jan, 2012) 023504},
  \href{http://arxiv.org/abs/1107.5427}{{\ttfamily arXiv:1107.5427}}.

\bibitem{bertacca:evolutionbias}
D.~Bertacca, R.~Maartens, A.~Raccanelli, and C.~Clarkson, ``Beyond the
  plane-parallel and Newtonian approach: wide-angle redshift distortions and
  convergence in general relativity'',
  \href{http://dx.doi.org/10.1088/1475-7516/2012/10/025}{{\em Journal of
  Cosmology and Astroparticle Physics} {\bfseries 2012} no.~10, (2012) 025},
  \href{http://arxiv.org/abs/1205.5221}{{\ttfamily arXiv:1205.5221}}.

\bibitem{nakamura:pbhmergerrate}
T.~Nakamura, M.~Sasaki, T.~Tanaka, and K.~S. Thorne, ``Gravitational Waves from
  Coalescing Black Hole MACHO Binaries'',
  \href{http://dx.doi.org/10.1086/310886}{{\em The Astrophysical Journal
  Letters} {\bfseries 487} no.~2, (1997) L139},
  \href{http://arxiv.org/abs/astro-ph/9708060}{{\ttfamily
  arXiv:astro-ph/9708060}}.

\bibitem{raidal:earlybinary}
M.~Raidal, V.~Vaskonen, and H.~Veerm\"ae, ``Gravitational waves from primordial
  black hole mergers'',
  \href{http://dx.doi.org/10.1088/1475-7516/2017/09/037}{{\em Journal of
  Cosmology and Astroparticle Physics} {\bfseries 2017} no.~09, (2017) 037},
  \href{http://arxiv.org/abs/1707.01480}{{\ttfamily arXiv:1707.01480}}.

\bibitem{tinker:halomassfunction}
J.~Tinker, A.~V. Kravtsov, A.~Klypin, K.~Abazajian, M.~Warren, G.~Yepes,
  S.~Gottl\"ober, and D.~E. Holz, ``Toward a Halo Mass Function for Precision
  Cosmology: The Limits of Universality'',
  \href{http://dx.doi.org/10.1086/591439}{{\em The Astrophysical Journal}
  {\bfseries 688} no.~2, (2008) 709},
  \href{http://arxiv.org/abs/0803.2706}{{\ttfamily arXiv:0803.2706}}.

\bibitem{mouri:pbhpairformation}
H.~Mouri and Y.~Taniguchi, ``Runaway Merging of Black Holes: Analytical
  Constraint on the Timescale'', \href{http://dx.doi.org/10.1086/339472}{{\em
  The Astrophysical Journal Letters} {\bfseries 566} no.~1, (2002) L17},
  \href{http://arxiv.org/abs/astro-ph/0201102}{{\ttfamily
  arXiv:astro-ph/0201102}}.

\bibitem{quinlan:pbhpairformation}
G.~D. Quinlan and S.~L. Shapiro, ``Dynamical evolution of dense clusters of
  compact stars'', \href{http://dx.doi.org/10.1086/167745}{{\em The
  Astrophysical Journal} {\bfseries 343} (Aug, 1989) 725--749}.

\bibitem{navarro:nfwhaloprofile}
J.~F. Navarro, C.~S. Frenk, and S.~D.~M. White, ``The Structure of Cold Dark
  Matter Halos'', \href{http://dx.doi.org/10.1086/177173}{{\em The
  Astrophysical Journal} {\bfseries 462} (May, 1996) 563},
  \href{http://arxiv.org/abs/astro-ph/9508025}{{\ttfamily
  arXiv:astro-ph/9508025}}.

\bibitem{prada:concentrationparameter}
F.~Prada, A.~A. Klypin, A.~J. Cuesta, J.~E. Betancort-Rijo, and J.~Primack,
  ``Halo concentrations in the standard $\Lambda$ cold dark matter cosmology'',
  \href{http://dx.doi.org/10.1111/j.1365-2966.2012.21007.x}{{\em Monthly
  Notices of the Royal Astronomical Society} {\bfseries 423} no.~4, (2012)
  3018--3030}, \href{http://arxiv.org/abs/1104.5130}{{\ttfamily
  arXiv:1104.5130}}.

\bibitem{ludlow:concentrationparameter}
A.~D. Ludlow, S.~Bose, R.~E. Angulo, L.~Wang, W.~A. Hellwing, J.~F. Navarro,
  S.~Cole, and C.~S. Frenk, ``The mass-concentration-redshift relation of cold
  and warm dark matter haloes'',
  \href{http://dx.doi.org/10.1093/mnras/stw1046}{{\em Monthly Notices of the
  Royal Astronomical Society} {\bfseries 460} no.~2, (2016) 1214--1232},
  \href{http://arxiv.org/abs/1601.02624}{{\ttfamily arXiv:1601.02624}}.

\bibitem{chisholm:pbhclusteringI}
J.~R. Chisholm, ``Clustering of primordial black holes: Basic results'',
  \href{http://dx.doi.org/10.1103/PhysRevD.73.083504}{{\em Phys. Rev. D}
  {\bfseries 73} (Apr, 2006) 083504},
  \href{http://arxiv.org/abs/astro-ph/0509141}{{\ttfamily
  arXiv:astro-ph/0509141}}.

\bibitem{chisholm:pbhclusteringII}
J.~R. Chisholm, ``Clustering of primordial black holes. II. Evolution of bound
  systems'', \href{http://dx.doi.org/10.1103/PhysRevD.84.124031}{{\em Phys.
  Rev. D} {\bfseries 84} (Dec, 2011) 124031},
  \href{http://arxiv.org/abs/1110.4402}{{\ttfamily arXiv:1110.4402}}.

\bibitem{alihaimoud:pbhclustering}
Y.~Ali-Ha\"{\i}moud, ``Correlation function of high-threshold peaks and
  application to the initial (non)clustering of primordial black holes'',
  \href{http://dx.doi.org/10.1103/PhysRevLett.121.081304}{{\em Phys. Rev.
  Lett.} {\bfseries 121} (Aug, 2018) 081304},
  \href{http://arxiv.org/abs/1805.05912}{{\ttfamily arXiv:1805.05912}}.

\bibitem{desjacques:pbhclustering}
V.~Desjacques and A.~Riotto, ``The Spatial Clustering of Primordial Black
  Holes'', \href{http://arxiv.org/abs/1806.10414}{{\ttfamily
  arXiv:1806.10414}}.

\bibitem{ballesteros:pbhclustering}
G.~Ballesteros, P.~D. Serpico, and M.~Taoso, ``On the merger rate of primordial
  black holes: effects of nearest neighbours distribution and clustering'',
  \href{http://arxiv.org/abs/1807.02084}{{\ttfamily arXiv:1807.02084}}.

\bibitem{mo:smallhalosbias}
H.~J. Mo and S.~D.~M. White, ``An analytic model for the spatial clustering of
  dark matter haloes'', \href{http://dx.doi.org/10.1093/mnras/282.2.347}{{\em
  Monthly Notices of the Royal Astronomical Society} {\bfseries 282} no.~2,
  (1996) 347--361}, \href{http://arxiv.org/abs/astro-ph/9512127}{{\ttfamily
  arXiv:astro-ph/9512127}}.

\bibitem{belczynsky:mergerrate}
K.~Belczynski, T.~Ryu, R.~Perna, E.~Berti, T.~L. Tanaka, and T.~Bulik, ``On the
  likelihood of detecting gravitational waves from Population III compact
  object binaries'', \href{http://dx.doi.org/10.1093/mnras/stx1759}{{\em
  Monthly Notices of the Royal Astronomical Society} {\bfseries 471} no.~4,
  (2017) 4702--4721}, \href{http://arxiv.org/abs/1612.01524}{{\ttfamily
  arXiv:1612.01524}}.

\bibitem{elbert:bhmergerrate}
O.~D. Elbert, J.~S. Bullock, and M.~Kaplinghat, ``Counting black holes: The
  cosmic stellar remnant population and implications for LIGO'',
  \href{http://dx.doi.org/10.1093/mnras/stx1959}{{\em Monthly Notices of the
  Royal Astronomical Society} {\bfseries 473} no.~1, (2018) 1186--1194},
  \href{http://arxiv.org/abs/1703.02551}{{\ttfamily arXiv:1703.02551}}.

\bibitem{Mapelli:merger_rate}
M.~Mapelli, N.~Giacobbo, E.~Ripamonti, and M.~Spera, ``The cosmic merger rate
  of stellar black hole binaries from the Illustris simulation'',
  \href{http://dx.doi.org/10.1093/mnras/stx2123}{{\em Monthly Notices of the
  Royal Astronomical Society} {\bfseries 472} no.~2, (2017) 2422--2435},
  \href{http://arxiv.org/abs/1708.05722}{{\ttfamily arXiv:1708.05722}}.

\bibitem{moore:gwsignaltonoise}
C.~J. Moore, R.~H. Cole, and C.~P.~L. Berry, ``Gravitational-wave sensitivity
  curves'', \href{http://dx.doi.org/10.1088/0264-9381/32/1/015014}{{\em
  Classical and Quantum Gravity} {\bfseries 32} no.~1, (2015) 015014},
  \href{http://arxiv.org/abs/1408.0740}{{\ttfamily arXiv:1408.0740}}.

\bibitem{vallisneri:gwsnr}
M.~Vallisneri and C.~R. Galley, ``Non-sky-averaged sensitivity curves for
  space-based gravitational-wave observatories'',
  \href{http://dx.doi.org/10.1088/0264-9381/29/12/124015}{{\em Classical and
  Quantum Gravity} {\bfseries 29} no.~12, (2012) 124015},
  \href{http://arxiv.org/abs/1201.3684}{{\ttfamily arXiv:1201.3684}}.

\bibitem{flanagan:gwsignaltonoise}
E.~E. Flanagan and S.~A. Hughes, ``Measuring gravitational waves from binary
  black hole coalescences. I. Signal to noise for inspiral, merger, and
  ringdown'', \href{http://dx.doi.org/10.1103/PhysRevD.57.4535}{{\em Phys. Rev.
  D} {\bfseries 57} (Apr, 1998) 4535--4565},
  \href{http://arxiv.org/abs/gr-qc/9701039}{{\ttfamily arXiv:gr-qc/9701039}}.

\bibitem{ajith:noisepsd}
P.~Ajith, ``Addressing the spin question in gravitational-wave searches:
  Waveform templates for inspiralling compact binaries with nonprecessing
  spins'', \href{http://dx.doi.org/10.1103/PhysRevD.84.084037}{{\em Phys. Rev.
  D} {\bfseries 84} (Oct, 2011) 084037},
  \href{http://arxiv.org/abs/1107.1267}{{\ttfamily arXiv:1107.1267}}.

\bibitem{huerta:etnoise}
E.~A. Huerta and J.~R. Gair, ``Intermediate-mass-ratio inspirals in the
  Einstein Telescope. I. Signal-to-noise ratio calculations'',
  \href{http://dx.doi.org/10.1103/PhysRevD.83.044020}{{\em Phys. Rev. D}
  {\bfseries 83} (Feb, 2011) 044020},
  \href{http://arxiv.org/abs/1009.1985}{{\ttfamily arXiv:1009.1985}}.

\bibitem{abbott:ligomergerrates}
The {\bfseries LIGO Scientific Collaboration and Virgo Collaboration}, B.~P.
  Abbott {\em et~al.}, ``Binary Black Hole Mergers in the First Advanced LIGO
  Observing Run'', \href{http://dx.doi.org/10.1103/PhysRevX.6.041015}{{\em
  Phys. Rev. X} {\bfseries 6} (Oct, 2016) 041015},
  \href{http://arxiv.org/abs/1606.04856}{{\ttfamily arXiv:1606.04856}}.

\bibitem{hayasaki:mergerrate}
K.~Hayasaki and A.~Loeb, ``Detection of Gravitational Wave Emission by
  Supermassive Black Hole Binaries Through Tidal Disruption Flares'',
  \href{http://dx.doi.org/10.1038/srep35629}{{\em Scientific Reports}
  {\bfseries 6} (2016) }, \href{http://arxiv.org/abs/1510.05760}{{\ttfamily
  arXiv:1510.05760}}.

\end{thebibliography}\endgroup
\bibliographystyle{utcaps}


\appendix
\section{Relativistic Number Counts}
\label{app:relativistic_number_counts}
In this appendix we explicitly report relativistic number counts effects, following the notation of Ref. \cite{didio:classgal}. The elements of equation \eqref{eq:numbercount_fluctuation} read as 
\begin{equation}
\begin{aligned}
\Delta_\ell^\mathrm{den}(k,z) &= b_X \delta(k,\tau_z) j_\ell,	\\
\Delta_\ell^\mathrm{vel}(k,z) &= \Delta_\ell^\mathrm{rsd}(k,z) + \Delta_\ell^\mathrm{dop}(k,z),	\\
\Delta_\ell^\mathrm{rsd}(k,z) &=  \frac{k}{\mathcal{H}}j''_\ell  V(k,\tau_z),	\\
\Delta_\ell^\mathrm{dop}(k,z) &= \left[(f^\mathrm{evo}_X-3)\frac{\mathcal{H}}{k}j_\ell + \left(\frac{\mathcal{H}'}{\mathcal{H}^2}+\frac{2-5s_X}{r(z)\mathcal{H}}+5s_X-f^\mathrm{evo}_X\right)j'_\ell \right]  V(k,\tau_z),	\\
\Delta_\ell^\mathrm{len}(k,z) &= \ell(\ell+1) \frac{2-5s_X}{2} \int_0^{r(z)} dr \frac{r(z)-r}{r(z) r} \left[\Phi(k,\tau_z)+\Psi(k,\tau_z)\right] j_\ell(kr),	\\
\Delta_\ell^\mathrm{gr}(k,z)  &= \left[\left(\frac{\mathcal{H}'}{\mathcal{H}^2}+\frac{2-5s_X}{r(z)\mathcal{H}}+5s_X-f^\mathrm{evo}_X+1\right)\Psi(k,\tau_z) + \left(-2+5s_X\right) \Phi(k,\tau_z) + \mathcal{H}^{-1}\Phi'(k,\tau_z)\right] j_\ell + \\
&+ \int_0^{r(z)} dr \frac{2-5s_X}{r(z)} \left[\Phi(k,\tau)+\Psi(k,\tau)\right]j_\ell(kr) , \\
&+ \int_0^{r(z)} dr \left(\frac{\mathcal{H}'}{\mathcal{H}^2}+\frac{2-5s_X}{r(z)\mathcal{H}}+5s_X-f^\mathrm{evo}_X\right)_{r(z)} \left[\Phi'(k,\tau)+\Psi'(k,\tau)\right] j_\ell(kr).
\end{aligned}
\end{equation}
According to the notation of Ref. \cite{didio:classgal}, $r$ is the conformal distance of on the light cone, $\tau=\tau_0-r$ is the conformal time, $\tau_z=\tau_0-r(z)$, $b_X$ is the bias parameter, $s_X$ is the magnification bias parameter, $f^\mathrm{evo}_X$ is the evolution bias parameter, Bessel functions and their derivatives $j_\ell$, $j'_\ell=\frac{dj_\ell}{dy}$, $j''_\ell=\frac{d^2j_\ell}{dy^2}$ are evaluated at $y=kr(z)$ unless explicitly stated, $\mathcal{H}$ is the conformal Hubble parameter, a prime $'$ indicates derivatives with respect to conformal time, $\delta$ is the density contrast in comoving gauge, $V$ is the peculiar velocity, $\Phi$ and $\Psi$ are Bardeen potentials.
 
The velocity term $\Delta_\ell^\mathrm{vel}(k,z)$ has been written in terms of the pure (Kaiser) redshift-space distortions term $\Delta_\ell^\mathrm{rsd}(k,z)$ and in term of Doppler contributions $\Delta_\ell^\mathrm{dop}(k,z)$. Notice that the magnification and evolution bias enter only in the Doppler term.

\end{document}